\documentclass[11pt]{article}

\usepackage[final]{acl}

\usepackage{times}
\usepackage{latexsym}
\usepackage{algorithm}
\usepackage{amsmath}
\usepackage{mathtools}
\usepackage{amsmath}
\usepackage{amssymb} 
\usepackage{amsfonts}
\usepackage{pifont}
\usepackage{multirow}
\usepackage{enumitem}
\usepackage{graphicx}
\usepackage{subcaption}
\usepackage{booktabs}
\usepackage{tabularx}  
\usepackage{enumitem}
\usepackage[most]{tcolorbox} 
\usepackage{listings} 
\usepackage{microtype}
\usepackage{algorithm}
\usepackage{algpseudocode}

\algnewcommand\algorithmicparfor{\textbf{parallel for}}
\algdef{S}[FOR]{ParFor}[1]{\algorithmicparfor\ #1\ \algorithmicdo}

\newtcblisting{AutoPrompt}[1]{
  enhanced,
  breakable,                
  listing only,             
  colback=white,
  colframe=gray!40!black,
  coltitle=white,
  fonttitle=\bfseries,
  title={#1},
  label={},
  attach boxed title to top left, 
  boxed title style={frame hidden}, 
  boxrule=0.5mm,
  sharp corners=south,
  listing options={
    basicstyle=\small\ttfamily,
    breaklines=true,        
    breakatwhitespace=true,
    columns=fullflexible,
    frame=none,
    aboveskip=0pt,
    belowskip=0pt,
    escapechar=|,           
    extendedchars=true,     
    literate={—}{{---}}1 
             {–}{{--}}1 
             {“}{{``}}1 
             {”}{{''}}1 
             {‘}{{`}}1
             {’}{{'}}1
             {→}{{$\rightarrow$}}1
             {←}{{$\leftarrow$}}1
             {…}{{\dots}}1
             {≠}{{$\neq$}}1
  }
}

\usepackage[T1]{fontenc}

\usepackage[utf8]{inputenc}

\usepackage{csquotes}

\usepackage{inconsolata}

\title{CogGen: A Cognitively Inspired Recursive Framework for Deep Research Report Generation}

\title{CogGen: A Cognitively Inspired Recursive Framework for Deep Research Report Generation}

\author{
  \textbf{Kuo Tian\textsuperscript{\rm 1,2}},
  \textbf{Pengfei Sun\textsuperscript{\rm 3}},
\textbf{Zhen Wu\textsuperscript{\rm 1,2}\thanks{Corresponding authors.}},
  \textbf{Junran Ding\textsuperscript{\rm 1,2}},
  \textbf{Xinyu Dai\textsuperscript{\rm 1,2}\footnotemark[1]}
\\
  \textsuperscript{\rm 1} National Key Laboratory for Novel Software Technology, Nanjing University, China\\
\textsuperscript{\rm 2} School of Artificial Intelligence, Nanjing University, China\\
  \textsuperscript{\rm 3} Nanjing Haodun Technology Development Co., Ltd.
\\
    \{tiank,jrding\}@smail.nju.edu.cn, chongqingspf@gmail.com,\{wuz,daixinyu\}@nju.edu.cn
}

\begin{document}
\maketitle
\begin{abstract}

The autonomous synthesis of deep research reports represents a critical frontier for Large Language Models (LLMs), demanding sophisticated information orchestration and non-linear narrative logic. Current approaches rely on rigid predefined linear workflows, which cause error accumulation, preclude global restructuring from subsequent insights, and ultimately limit in-depth multimodal fusion and report quality. We propose \textbf{CogGen}, a \textbf{Cog}nitively inspired recursive framework for deep research report \textbf{Gen}eration. Leveraging a Hierarchical Recursive Architecture to simulate cognitive writing, CogGen enables flexible planning and global restructuring. To extend this recursivity to multimodal content, we introduce Abstract Visual Representation (AVR): a concise intent-driven language that iteratively refines visual-text layouts without pixel-level regeneration overhead. We further present \textbf{CLEF}, a \textbf{C}ognitive \textbf{L}oad \textbf{E}valuation \textbf{F}ramework, and curate a new benchmark from \textit{Our World in Data} (OWID). Extensive experiments show CogGen achieves state-of-the-art results among open-source systems, generating reports comparable to professional analysts’ outputs and surpassing Gemini Deep Research. Our code and dataset are available at \url{https://github.com/NJUNLP/CogGen}.
\end{abstract}

\section{Introduction}
\label{sec:Introduction}

Driven by advancements in reasoning and tool-use capabilities~\cite{OpenAI2025O1SystemCard,IntroducingClaude35,deepseek-aiDeepSeekR1IncentivizingReasoning2025}, Large Language Models (LLMs) have demonstrated the potential to autonomously synthesize structured deep research reports~\cite{zhangWebSearchAgentic2025,liSearcho1AgenticSearchEnhanced2025}. However, bridging the gap between automated generation and expert-level analytical writing remains a formidable challenge~\cite{zhengDeepResearcherScalingDeep2025,duDeepResearchBenchComprehensive2025a}. Expert report writing is not a mere assembly of retrieved facts; it is a sophisticated cognitive process characterized by recursive refinement and the seamless integration of heterogeneous evidence.

\begin{figure}[t]
\centering
\includegraphics[width=0.9\columnwidth]{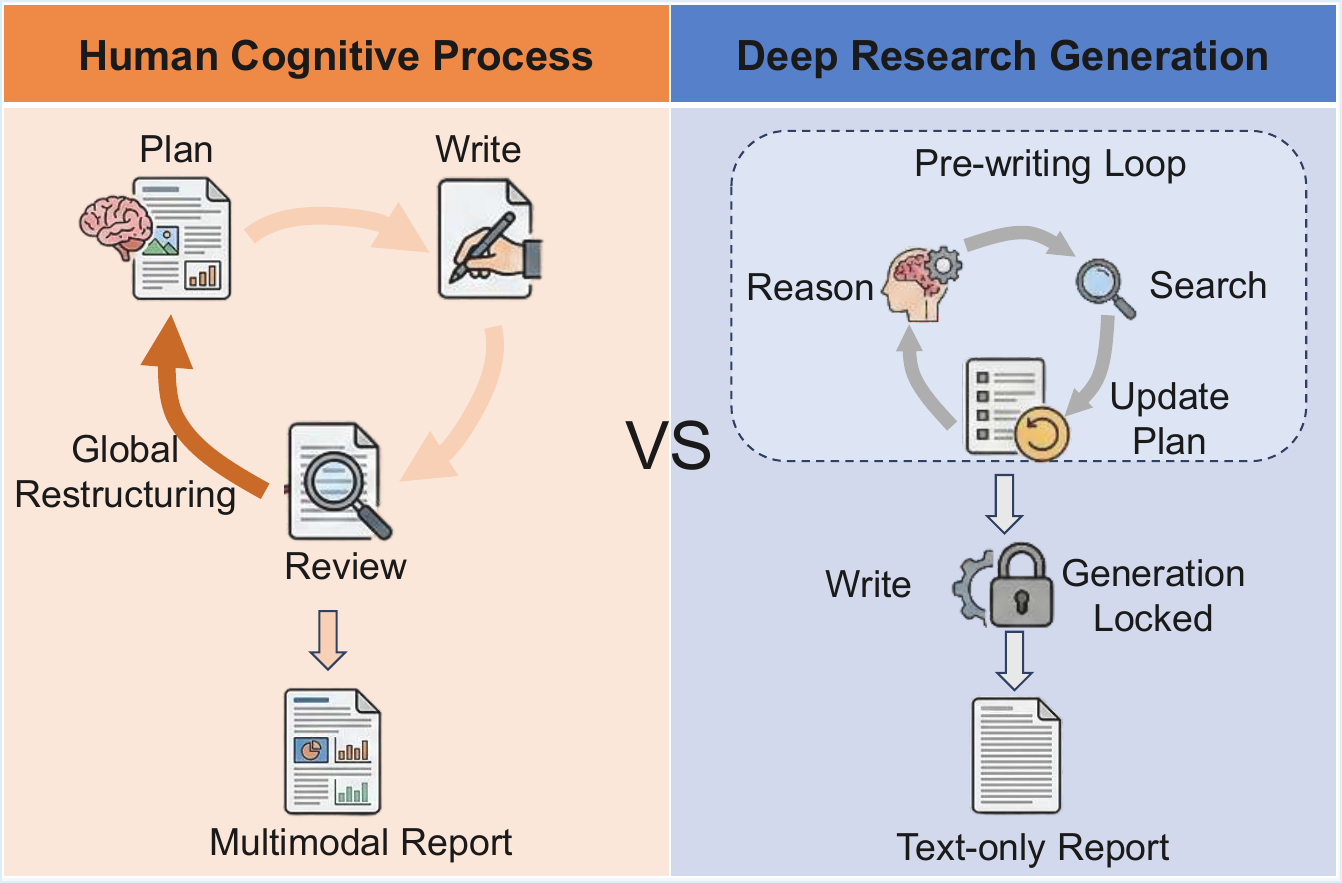} 
\caption{Comparison of report writing paradigms.  
The Human Cognitive Process (left) adopts a recursive ``plan-write-review'' loop that supports global restructuring throughout the writing process.
In contrast, the Deep Research Report Generation (right) relies on a linear workflow, where once the preceding content is generated, it cannot be modified in reverse and limits the generation of subsequent sections.
}
\label{fig1}
\end{figure}

Existing deep research report generation paradigms primarily fall into two architectural categories: single-agent systems that integrate reasoning models with complex tool invocation~\cite{Google2025DeepResearch,OpenAI2025DeepResearch} and multi-agent frameworks that incorporate role-playing coupled with feedback mechanisms~\cite{shaoAssistingWritingWikipedialike2024a,jiangUnknownUnknownsEngaged2024,wangAutoSurveyLargeLanguage2024}. Despite being well-designed, both structures typically follow a linear, predefined execution workflow. Once a plan is drafted, the generation follows a forward-only path, making it difficult for existing agent frameworks to perform the ``backward restructuring'' necessary when downstream discoveries invalidate earlier organizational logic~\cite{xuComprehensiveSurveyDeep2025}. As illustrated in Figure \ref{fig1}, this linear rigidity stands in stark contrast to the human cognitive writing process, which functions as an inherently non-linear, recursive mechanism of exploration.

Furthermore, true deep research necessitates the integration of quantitative visual evidence (e.g., charts) to substantiate qualitative claims. However, current multimodal efforts typically generate these elements separately from the text~\cite{shiCalliopeAutomaticVisual2021, yangFinRobotOpenSourceAI2024a}. This asynchronous generation creates a superficial relationship between text and image, where a chart might be redundant to the text or lack the specific data granularity mentioned in the narrative. This forces the reader to manually bridge the gap between abstract descriptions and visual data, leading to a fragmented cognitive experience where the visual acts as a mere illustration rather than a synergistic argument.

To address these issues, we propose \textbf{CogGen}, a cognitively inspired multi-agent framework emulating the recursive nature of expert writing. Drawing on the Cognitive Process Theory of Writing~\cite{flowerCognitiveProcessTheory1981,hayesNewFrameworkUnderstanding1996}, we introduce a \textbf{Hierarchical Recursive Architecture}. This architecture comprises a \textit{Macro-Cognitive Loop} for global logic orchestration and a \textit{Micro-Cognitive Cycle} for autonomous intra-section refinement. By enabling agents to dynamically pause, review, and restructure the global plan based on emerging information, CogGen transcends the ``linear lock-in'' of traditional paradigm, allowing for a fluid and logically coherent narrative evolution.

Beyond structural logic, CogGen addresses the multimodal integration gap through the lens of Cognitive Offloading~\cite{riskoCognitiveOffloading2016}. Research suggests that expert writers often decouple high-level content planning from low-level visual rendering to mitigate dual-task interference. Consistent with this behavior, we introduce an \textbf{Abstract Visual Representation (AVR)}. By abstracting verbose visualization specifications into a compact intermediate representation, this schema allows the agent to treat visual elements as mutable semantic tokens while offloads the final visualization to specialized rendering agents. This enables the synchronous iteration of narrative and visual plans with minimal cognitive load, ensuring that charts and text achieve a high degree of synergy rather than mere alignment.


To rigorously evaluate the quality of synthesized reports, we propose the \textbf{Cognitive Load Evaluation Framework (CLEF)}. Moving beyond surface-level n-gram metrics, CLEF is grounded in cognitive load theory~\cite{sweller1994cognitive}, assessing reports across five dimensions: Organization, Depth, and Relevance, Alignment, Synergy. We benchmark CogGen on a newly curated dataset from \textit{Our World in Data} (OWID) and the \textit{WildSeek} benchmark. Experimental results demonstrate that CogGen significantly outperforms state-of-the-art open-source frameworks. Notably, CogGen-generated reports achieved parity with human expert benchmarks on OWID and surpassed references from Gemini Deep Research on WildSeek.

Our primary contributions are as follows:
\begin{itemize}
    \item \textbf{Framework:} We propose novel CogGen, a Hierarchical Recursive Framework that operationalizes cognitive writing theories to enable non-linear, global logic restructuring in deep research reports generation.
    \item \textbf{Mechanism:} We introduce an Abstract Visual Representation rooted in cognitive offloading theory, facilitating the deep semantic integration of text and visual evidence.
    \item \textbf{Evaluation:} We present CLEF, a cognitive theory-driven evaluation framework, and release a high-quality benchmark based on OWID to facilitate future research in deep research agents.
\end{itemize}

\begin{figure*}[t]
\centering
\includegraphics[width=0.9\textwidth]{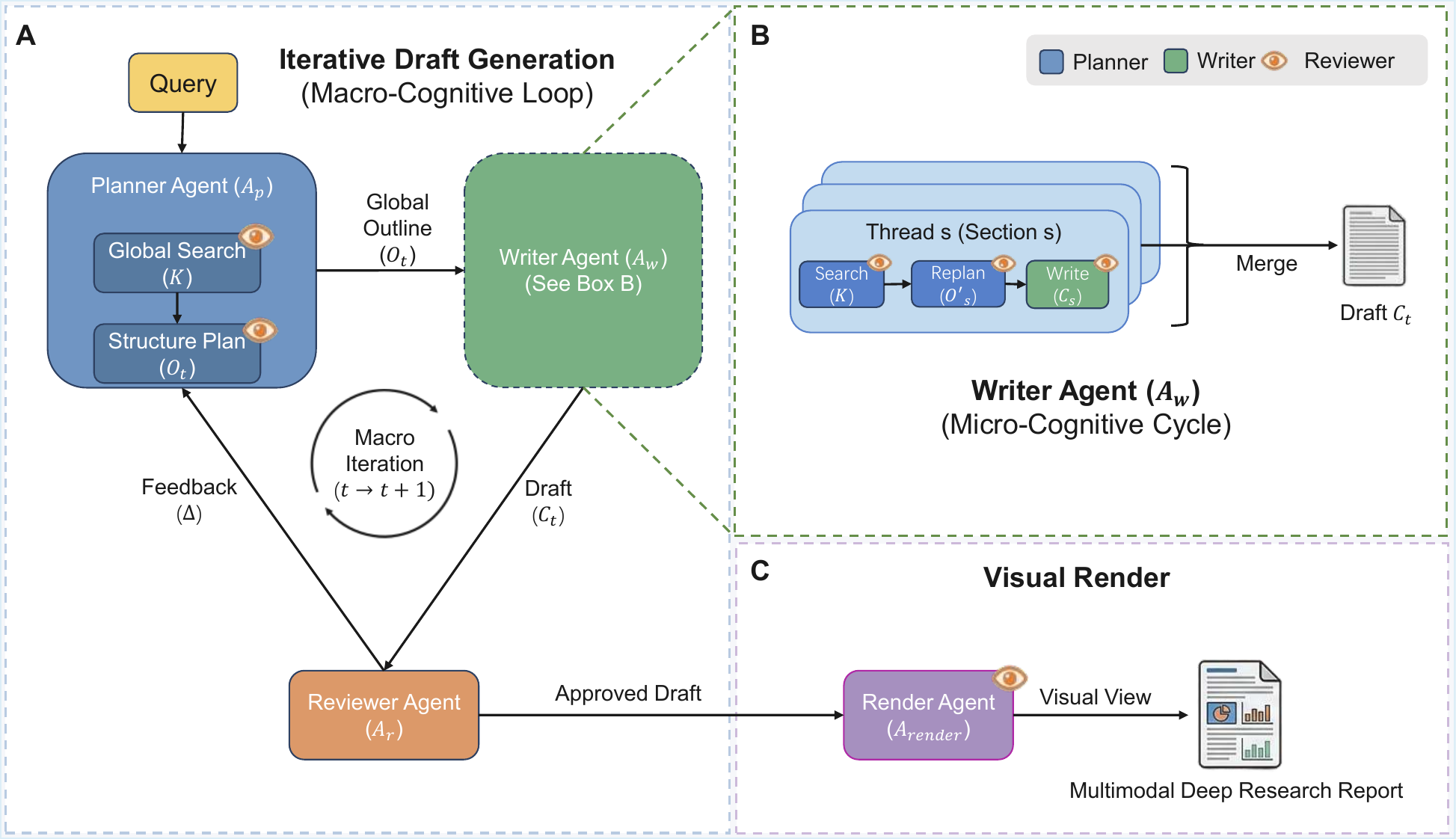} 
\caption{\label{fig:framework}
    Overview of the CogGen framework. Components marked with an eye icon indicate operations strictly monitored by the Reviewer Agent ($A_r$) to enable feedback-driven iteration.
    \textbf{(A) Macro-Cognitive Loop:} A global iterative process consisting of three phases. The Planner Agent ($A_p$) generates the outline ($O^{(t)}$), the Writer Agent ($A_w$) produces the draft ($C^{(t)}$), and the Reviewer Agent ($A_r$) evaluates the complete draft to generate feedback ($\Delta$) for the next iteration.
    \textbf{(B) Micro-Cognitive Cycle:} Within the Writer Agent ($A_w$), multiple threads execute monitored ``Search--Replan--Write'' cycles to generate section drafts ($C_s$), which are finally merged into the draft ($C_t$).
    \textbf{(C) Visual Rendering:} In the Execution phase, the Render Agent ($A_{\text{render}}$) translates the approved draft into a visual view, operating under the Reviewer's supervision to ensure alignment with the visual specifications.
}
\label{fig2}
\end{figure*}

\section{Related Work}
\label{sec:related work}


\subsection{Agentic Report Generation}
Prior automated report generation primarily relied on domain-specific fixed workflows~\cite{wangAutoSurveyLargeLanguage2024,ghafarollahiSciAgentsAutomatingScientific2025,zhangLLMbasedMultiagentPoetry2024,pichlmairDramaEngineFramework2024,huotAgentsRoomNarrative2025}, whose performance was constrained by predefined linear processes. Concurrent works attempt to mitigate this via dynamic retrieval; however, PAGER~\cite{liStructuredKnowledgeRepresentation2026} targets QA tasks rather than long-form generation, and Mind2Report~\cite{chengMind2ReportCognitiveDeep2026} retains a unidirectional serial workflow lacking global restructuring. To address complex tasks, frameworks like WriteHere~\cite{xiongOutliningHeterogeneousRecursive2025} and ReCode~\cite{yuReCodeUnifyPlan2026} introduce recursive decomposition. Yet, they remain essentially forward-generation methods unable to retroactively resolve structural disruptions. Similarly, while ARCS~\cite{bhattaraiARCSAgenticRetrievalAugmented2025} utilizes execution-repair loops, its global granularity scales poorly to comprehensive reports. Other studies enhance planning via role-playing~\cite{shaoAssistingWritingWikipedialike2024a,jiangUnknownUnknownsEngaged2024}, failing to address the disconnect between writing and planning. Furthermore, despite recent advances in verification-centric evaluations like DEER~\cite{hanDEERComprehensiveReliable2025}, even state-of-the-art commercial models (e.g., OpenAI~\cite{OpenAI2025DeepResearch} and Gemini Deep Research~\cite{Google2025DeepResearch}) remain limited by fixed frameworks during their writing execution stage. In contrast, CogGen proposes a recursive outline modification mechanism (Global Restructuring) to iteratively refine both historical and future content contextually.

\subsection{Multimodal Report Generation}
\label{sec:mm_report}

Early multimodal report generation primarily relied on domain-specific frameworks~\cite{shiCalliopeAutomaticVisual2021, yangFinRobotOpenSourceAI2024a}, adopting a sequential slot-filling strategy to generate text and visuals independently. Recent works such as Multimodal DeepResearcher~\cite{yangMultimodalDeepResearcherGenerating2025} enabled open-domain multimodal generation by introducing visual description languages~\cite{2017-vega-lite} and embedding chart generation into linear workflows. However, they are essentially loose combinations of text and visual generation without in-depth collaborative optimization. In contrast, CogGen introduces the Abstract Visual Representation and shifts the objective from visual fidelity to the characterization of visual semantic intent, achieving semantic-level collaborative planning and iterative optimization of both textual and visual content.

\section{Methodology}
\label{sec:Methodology}

\subsection{Framework Overview}
\label{sec:Framework_Overview}

To overcome the linear constraints discussed in Section~\ref{sec:Introduction}, CogGen implements Hierarchical Recursive Architecture (Figure~\ref{fig:framework}). Instead of a static chain, this design treats the generation plan as a mutable object, enabling dynamic, non-linear transitions across planning, writing, and reviewing phases.

Formally, we model report generation as a mapping from a user query $Q$ to a multimodal deep research report $R$, denoted as $R = \text{CogGen}(Q)$. The process is collaboratively executed by three peer cognitive agents (Figure~\ref{fig:framework}A):

\begin{itemize}
    \item Planner Agent ($A_p$): Responsible for information retrieval and structural planning. Its function is formalized as a mapping $\mathcal{O}, K = A_p(Q, \mathcal{H})$, where $\mathcal{H}$ represents the interaction history and feedback state. $Q$ is the user query, $\mathcal{O}$ is the writing outline, and $K$ is the knowledge base formed by information retrieved during outline generation.
    \item Writer Agent ($A_w$): Responsible for text composition and the definition of visual intent. Its function is formalized as $C = A_w(\mathcal{O}, K)$, where $C$ represents the draft with the abstract vision representations (AVRs) generated by the writing agent.
    \item Reviewer Agent ($A_r$): An integrated evaluation engine with dual functions of real-time monitoring and post-hoc assessment. By outputting feedback signals $\Delta$, this agent achieves two core objectives: ensuring the generation process adheres to preset constraints under monitoring mode, and optimizing content quality under reviewing mode.
\end{itemize}

Unlike traditional linear chain structures~\cite{shaoAssistingWritingWikipedialike2024a,yangWikiAutoGenMultiModalWikipediaStyle2025}, this collaborative agent triad supports recursive operations at both the macro (global report) and micro (local section) granularities, as illustrated in parts A and B of Figure~\ref{fig:framework}, ensuring generation quality through immediate review mechanisms.

\subsection{Macro-Cognitive Loop}
\label{sec:Macro_Mechanism}
The core engine of CogGen is designed to enable Global Restructuring. To address the rigidity of linear workflows, where the generated preceding content cannot be reconstructed in reverse~\cite{xuComprehensiveSurveyDeep2025}, CogGen utilizes a Macro-Cognitive Loop to implement recursive optimization.

This mechanism empowers the system to perform backward restructuring: it allows agents to retroactively refine the global outline ($\mathcal{O}$) and previously generated drafts based on downstream discoveries. This ensures that the final report maintains global logical coherence rather than being a linear accumulation of sub-tasks. In the loop shown in Figure~\ref{fig:framework}, $t$ represents the iteration round.

\subsubsection{Iterative Global Planning}
The process begins with macro planning. First, the Planner Agent ($A_p$) 
performs a breadth-first retrieval to construct the initial knowledge base $K$ and a report blueprint, denoted formally as the outline $\mathcal{O}^{(0)}$. This corresponds to the initial state where history is empty ($\mathcal{H}=\emptyset$):
\begin{equation}
    \mathcal{O}^{(0)}, K = A_p(Q, \emptyset)
\end{equation}

To support parallel generation (Section~\ref{sec:Micro_Mechanism}), $K$ adopts a hierarchical architecture: a shared global snapshot provides common context to all generation threads, while section-specific evidence retrieved during micro-cycles is maintained in thread-local caches. This design prevents irrelevant noise from propagating across unrelated chapters while ensuring each thread retains the targeted evidence required for deep synthesis. A formal specification of this protocol is provided in Appendix~\ref{sec:parallel_formal}.

In subsequent rounds ($t>0$), $A_p$ refines the structure based on the feedback signal $\Delta^{(t)}$ derived from the previous draft $C^{(t)}$. This constitutes the ``Macro-Cognitive Loop'' (Figure~\ref{fig:framework}A), enabling retroactive adjustments to global logic:
\begin{equation}
    \mathcal{O}^{(t+1)} = A_p(Q, \{\mathcal{O}^{(t)}, \Delta^{(t)}\} \mid K)
\end{equation}
This recursive update ensures that the narrative structure and visual planning co-evolve, preventing the logical inconsistencies typical of static planning approaches.

\begin{table}[t]
\centering
\small 
\begin{tabular}{@{}l@{}}
\toprule
\textbf{Structure of Abstract Visual Representation ($P_{\text{vis}}$)} \\ \midrule
\texttt{[DATA\_VISUALIZATION]} \\
\quad \textbf{Title:} Adoption of Key AI Technologies in Michelin... \\
\quad \textbf{Chart\_Type:} Bar Chart \\
\quad \textbf{X\_Axis:} Types of AI Technology (Chatbots, Robotics... \\
\quad \textbf{Y\_Axis:} Estimated Adoption Level in Restaurants... \\
\quad \textbf{Data\_Source:} \textless{}ref:1003\textgreater \\
\quad \textbf{Purpose:} To visually compare the adoption rates... \\
\texttt{[/DATA\_VISUALIZATION]} \\ \bottomrule
\end{tabular}
\caption{An instantiation of the Abstract Visual Representation (AVR). The Writer generates this structured semantic representation instead of executable code, decoupling reasoning from rendering.}
\label{tab:viz_placeholder}
\end{table}

\subsubsection{Parallel Multimodal Content Writing}
To improve report synthesis efficiency, CogGen generates multiple sections in parallel (details are shown in Section~\ref{sec:Micro_Mechanism}). Specifically, the Writer Agent $A_w$ generates a unified draft $C^{(t)}$ based on the global outline. To ensure parallel consistency, the generation of each section strictly follows the constraints of the global outline $\mathcal{O}^{(t)}$:
\begin{equation}
    C^{(t)} = \{ A_w(o_s, \mathcal{O}^{(t)}, K) \mid \forall o_s \in \mathcal{O}^{(t)} \}
\end{equation}
By using the global structure $\mathcal{O}^{(t)}$ as a constraint, all parallel generation threads maintain consistency with the overall logic of the report. The draft $C^{(t)}$ contains both textual content and AVRs ($P_{\text{vis}}$). These vision representations carry complete visualization intents (shown in Table~\ref{tab:viz_placeholder}) but use a highly structured description to reduce cognitive load.

\subsubsection{Global Review}
The Reviewer Agent $A_r$ conducts a comprehensive evaluation of the current draft $C^{(t)}$ and outputs a feedback signal $\Delta^{(t)}$. This signal contains optimization suggestions for the current outline based on the newly generated draft. The feedback signal $\Delta^{(t)}$ serves as the input for the next round of planning, thereby driving the co-evolution of text and visual content through the recursive loop.

To enforce stability, CogGen incorporates a strict monotonic improvement constraint. Rather than relying on open-ended refinement, the system accepts a global update only when the Reviewer Agent validates a distinct increase in report quality. By rejecting changes that fail to meet this evaluation threshold, the architecture is designed to suppress infinite oscillation and drive the draft towards a local optimum relative to the reviewer's criteria. 
Appendix~\ref{sec:Appendix_Convergence} provides a \textit{theoretical analysis} of the convergence properties of this mechanism, modeling CogGen as a bounded state-space search with empirically validated stability.

\subsection{Micro-Cognitive Cycle}
\label{sec:Micro_Mechanism}

While the macro mechanism maintains global coherence, the detailed content generation is handled via parallelized micro-cycles. As illustrated in Part B of Figure~\ref{fig:framework}, the Writer Agent does not generate linearly; instead, it orchestrates multiple independent threads in parallel, recursively invoking the capabilities of the Planner and Reviewer Agents.

\textbf{Recursive Execution Flow.}
Consistent with the workflow depicted in Figure~\ref{fig:framework}, each section generation thread ($Thread_s$) executes a recursive ``Search--Replan--Write'' process:
\begin{itemize}
\item Search and Replan: The thread temporarily re-engages the Planner Agent to perform targeted retrieval and, if necessary, adaptively adjusts the section's internal outline based on retrieved evidence.
\item Write: The Writer Agent then composes the section text based on the retrieved evidence and refined outline.
\item Review: The search, replan, and write processes are continuously monitored by the Reviewer Agent. Any intermediate state or final content that deviates from the requirements triggers an immediate correction loop, ensuring that errors are caught and resolved before propagating to the next stage.
\end{itemize}

\textbf{Parallelism and Deferred Update.}
Integrating retroactive revision into a serial workflow introduces critical stability issue we term Contextual Oscillation: correcting an upstream section (e.g., Sec 1) to align with a downstream discovery (e.g., Sec 5) invalidates the intermediate context. 
Without a global perspective, the model performs myopic corrections—fixing Sec 1 creates new inconsistencies with Sec 5, triggering a recursive modification loop between chapters~\cite{huangLargeLanguageModels2024}. 
Since the draft is incomplete during this serial process, the agent lacks the holistic view required to resolve these cross-section conflicts, leading to non-convergence.

To break the issue from recursive loops inherent in serial revision, CogGen employs a parallel architecture with a \textit{Deferred Update Policy}: parallel micro-cycles operate as read-only observers of the global outline $\mathcal{O}^{(t)}$, with section-specific retrieval confined to thread-local caches. Cross-section conflicts are not resolved locally but deferred to the Reviewer Agent $A_r$, which serves as the sole arbitrator during macro-cycle transitions (Appendix~\ref{sec:parallel_formal}). Under this policy, $A_r$ aggregates all cross-section conflicts into a global feedback signal $\Delta^{(t)}$.
\begin{equation}
    \Delta^{(t)} \leftarrow A_r(\mathcal{C}^{(t)}, \mathcal{O}^{(t)})
\end{equation}
This signal provides high-level guidance for the subsequent replanning phase ($\mathcal{O}^{(t+1)}$). By resolving conflicts at the global outline level rather than the local text level, CogGen ensures that structural adjustments are coherently propagated across all dependent chapters. A theoretical analysis of convergence properties is provided in Appendix~\ref{sec:Appendix_Convergence}, with empirical validation in Appendix~\ref{app:data_analysis}.

\subsection{Visual Rendering Engine}
\label{sec:Execution_Mechanism}

To efficiently handle multimodal fusion, we operationalize the Cognitive Offloading strategy proposed in Section~\ref{sec:Introduction}. Instead of disrupting the reasoning flow with complex code generation~\cite{sweller1994cognitive}, the Writer Agent ($A_w$) employs an Abstract Visual Representation mechanism. It focuses solely on the visual intent ($P_{\text{vis}}$)---describing data points and chart types without implementation details (as shown in Table~\ref{tab:viz_placeholder}, detailed in appendix~\ref{sec:appendix_rendering}).

This design contrasts with the Formal Description of Visualization (FDV) adopted by prior work~\cite{yangWikiAutoGenMultiModalWikipediaStyle2025}: while FDV requires the Writer to simultaneously specify visual styling, layout, and data, AVR captures only semantic intent (\textit{what} to show and \textit{why}), offloading visual design decisions to a dedicated Render Agent. This separation of concerns frees the Writer's cognitive resources for narrative reasoning and provides a natural insertion point for post-rendering data verification. A quantitative comparison is presented in Section~\ref{sec:avr_efficacy}.

Subsequently, the Renderer Agent ($A_{\text{render}}$) acts as a code interpreter, translating these semantic intents into executable syntax ($P_{\text{syn}}$) using libraries such as ECharts~\cite{liEChartsDeclarativeFramework2018} or Mermaid~\cite{Sveidqvist_Mermaid_Generate_diagrams_2014}. This generation process includes a syntax validation check to ensure executability before rendering the final style-consistent visual assets ($V$) in a headless browser. The pipeline is formalized as:

\begin{equation}
\begin{aligned}
    P_{\text{syn}} &= A_{\text{render}}(P_{\text{vis}})\\
    V &= \text{Browser}(P_{\text{syn}}) 
\end{aligned}
\end{equation}

This two-stage rendering scheme reduces the cognitive load during the writing and planning phases by decoupling the visual planning and generation stage from the rendering stage.

\begin{table}[t]
\centering
\small
\begin{tabularx}{\columnwidth}{@{} l X @{}}
\toprule
\textbf{Dimension} & \textbf{Evaluation Focus} \\
\midrule
D1: Organization & Hierarchical structure and navigation \\
D2: Depth & Causal explanations and schema construction \\
D3: Relevance & Appropriate complexity and coherence \\
D4: Alignment & Spatial/semantic integration of visuals and text \\
D5: Synergy & Information complementarity beyond text \\
\bottomrule
\end{tabularx}
\caption{Overview of CLEF's five evaluation dimensions grounded in Cognitive Load Theory.}
\label{tab:clef_overview}
\end{table}

\begin{table*}[t]
\centering
\small
\setlength{\tabcolsep}{5pt}
\begin{tabular}{l|ccccc|c}
\toprule
\textbf{Model} & \textbf{Organization} & \textbf{Depth} & \textbf{Relevance} & \textbf{Alignment} & \textbf{Synergy} & \textbf{Avg. Score} \\ \midrule
\multicolumn{7}{c}{\textbf{\textit{Dataset I: OWID (High-Density Multimodal Reports)}}} \\ \midrule
Human Gold-Standard (Ref) & \textbf{0.4986} & 0.5000 & \underline{0.5000} & \textbf{0.5000} & \textbf{0.5000} & \textbf{0.4997} \\
STORM & 0.4253 & 0.4443 & 0.3986 & 0.1675 & 0.1667 & 0.3205 \\
Co-STORM & 0.4132 & 0.4261 & 0.4281 & 0.1794 & 0.1667 & 0.3227 \\
Multimodal DeepResearcher & 0.3768 & 0.4293 & 0.3508 & 0.1819 & 0.1700 & 0.3018 \\
WriteHere & 0.4912 & \underline{0.5503} & 0.4936 & 0.3846 & 0.3312 & 0.4502 \\ \midrule
\textbf{CogGen (Ours)} & \underline{0.4972} & \textbf{0.5813} & \textbf{0.5042} & \underline{0.4806} & \underline{0.4326} & \underline{0.4992} \\ \midrule
\multicolumn{7}{c}{\textbf{\textit{Dataset II: WildSeek (Text-Centric Complex Queries)}}} \\ \midrule
Gemini Deep Research (Ref) & 0.5000 & \textbf{0.5000} & 0.5000 & \underline{0.5000} & \underline{0.5000} & \underline{0.5000} \\
STORM & 0.4375 & 0.4097 & 0.4472 & 0.1903 & 0.1908 & 0.3351 \\
Co-STORM & 0.3993 & 0.3695 & 0.4270 & 0.1943 & 0.1834 & 0.3147 \\
Multimodal DeepResearcher & 0.3819 & 0.3740 & 0.3695 & 0.2076 & 0.2183 & 0.3103 \\
WriteHere & \underline{0.5243} & 0.4931 & \underline{0.5271} & 0.4738 & 0.4497 & 0.4936 \\ \midrule
\textbf{CogGen (Ours)} & \textbf{0.5389} & \underline{0.5000} & \textbf{0.5334} & \textbf{0.5544} & \textbf{0.5437} & \textbf{0.5341} \\ \bottomrule
\end{tabular}
\caption{\textbf{Main Results on Multimodal Report Generation.} Scores represent the Relative Advantage Score ($R$) based on pairwise comparison against the Reference (Ref). A score of 0.5000 indicates parity with the reference; values > 0.5 indicate the model outperforms the reference. CogGen achieves comparable performance to Human Experts in overall quality (Avg. Score) on the data-intensive OWID dataset, driven by superior Depth and Relevance, and outperforms Gemini Deep Research on the text-centric WildSeek dataset. The best results are highlighted in \textbf{bold}, and the second-best are \underline{underlined}.}
\label{tab:main_results}
\end{table*}

\section{Experimental Setup}
\label{sec:setup}

In this section, we detail the experimental configuration used to evaluate CogGen's performance. We first introduce the two datasets used for evaluating report generation capabilities, then define the baseline models used for comparison, and finally elaborate on our proposed evaluation metrics based on cognitive load theory~\cite{sweller1994cognitive}.

\subsection{Datasets}
To comprehensively evaluate the model's capability in generating high-quality deep research reports, we employ a hybrid evaluation strategy combining a self-constructed dataset with an established benchmark. Given the scarcity of existing datasets containing professional-grade reports with rich data visualizations, we curated the OWID dataset to serve as a gold standard for complex multimodal generation. Complementarily, we adopt WildSeek, a standard dataset from prior work~\cite{jiangUnknownUnknownsEngaged2024}, to assess the model's robustness in handling diverse user intents within open-domain scenarios.

\paragraph{OWID.}
This dataset contains 40 research reports collected from the Our World in Data (OWID) website. Written by professional analysts, these reports feature substantial data density and logical depth, and include rich data visualizations. Detailed procedures for dataset construction and preprocessing are provided in Appendix \ref{app:dataset}. We use these reports as the Human Gold-Standard to evaluate the model's ability to generate comprehensive and high-quality multimodal content.

\paragraph{WildSeek.}
WildSeek~\cite{jiangUnknownUnknownsEngaged2024} was originally a standard dataset for evaluating pure text report generation. To adapt to the objectives of this study, we manually selected 20 queries with clear multimodal generation tendencies (e.g., questions requiring trend comparison or distribution display) to test the robustness of the model in generating illustrated reports in open-domain scenarios.

\subsection{Baselines}
We benchmark CogGen against a comprehensive set of baselines representing distinct generation paradigms: (1) STORM~\cite{shaoAssistingWritingWikipedialike2024a} and Co-STORM~\cite{jiangUnknownUnknownsEngaged2024}, the standard baselines for multi-perspective QA and collaborative writing; (2) WriteHere~\cite{xiongOutliningHeterogeneousRecursive2025}, the current state-of-the-art open-source model; and (3) Multimodal DeepResearcher~\cite{yangMultimodalDeepResearcherGenerating2025}, which represents linear multimodal generation workflows.

\noindent\textbf{Reference Standards.} For the OWID dataset, human-authored reports serve as the gold standard. For the WildSeek dataset, which lacks human ground truth, we adhere to established protocols~\cite{duDeepResearchBenchComprehensive2025a} by employing outputs from Gemini Deep Research~\cite{Google2025DeepResearch} as a commercial reference anchor for scoring.

\subsection{Metrics: Cognitive Load Evaluation}
Existing evaluation metrics present significant limitations when applied to multimodal deep research reports. Mechanical metrics~\cite{papineni2002bleu,lin2004rouge} focus on textual n-gram overlap, failing to capture the quality of text and visual elements from semantics. Similarly, while standard LLM-as-a-Judge approaches~\cite{zheng2023judging} assess general semantic quality, they lack a theoretical grounding to evaluate the cognitive synergy between modalities. Specifically, whether visual aids reduce the reader's mental effort. To bridge these gaps, we propose the Cognitive Load Evaluation Framework (CLEF), grounded in Cognitive Load Theory~\cite{sweller1994cognitive} and Mayer's Cognitive Theory of Multimedia Learning~\cite{mayer2005cognitive}.

CLEF operationalizes 11 of Mayer's 14 multimedia principles into five orthogonal evaluation dimensions. Table~\ref{tab:clef_overview} provides an overview of the five dimensions. These dimensions are organized into two categories: \textit{Control Dimensions} (D1-D3) ensuring general content quality, and \textit{Core Dimensions} (D4-D5) focusing on multimodal integration quality. Three CTML principles (Modality, Temporal Contiguity, Voice) are excluded as they specifically address dynamic multimedia and are not applicable to static text-visual reports. Notably, our evaluation framework explicitly classifies tables as visual modalities. This decision is grounded in Cognitive Load Theory, which posits that tabular organization—like graphical elements—significantly mitigates cognitive load. While CLEF operationalizes established cognitive principles into measurable dimensions rather than directly measuring reader behavior (e.g., subjective workload), its validity is supported by high consistency with human expert judgments (Section~\ref{sec:human_eval}) and robustness across multiple evaluation models (Appendix~\ref{app:detailed_eval}).

Following recent best practices~\cite{duDeepResearchBenchComprehensive2025a,krumdick2025no}, we employ pairwise comparison using GPT-5~\cite{IntroducingGPT52025} as the evaluator. For each dimension, we calculate the Relative Advantage Score ($R \in [0,1]$), where $R > 0.5$ indicates the model outperforms the baseline in enhancing understanding or reducing cognitive burden. Complete theoretical foundations, detailed dimension definitions, scoring mechanisms, and validation results are provided in Appendix~\ref{app:clef}.

\subsection{Implementation Details}
CogGen is implemented using a multi-agent architecture. The search tool utilizes GPT-4.1-Mini~\cite{OpenAI2025GPT41} for cost-effective query expansion, while the Planner, Writer, Reviewer and Render Agents utilize GPT-4.1 to ensure reasoning depth. To balance generation diversity and stability, we set the temperature to 0.5 for all agents. The external retrieval tool is the Tavily Search~\cite{tavily2025searchapi}. \textbf{Notably}, for fair comparison, the backbone LLM of baselines were unified to GPT-4.1, and the retrieval tool was unified to Tavily Search.

\begin{table*}[t]
  \centering
  \small
  \setlength{\tabcolsep}{3.5pt}
  \begin{tabular}{l|cc|ccccc|c}
  \toprule
  \multirow{2}{*}{\textbf{Method Variants}} & \multicolumn{2}{c|}{\textbf{Core Mechanisms}} & \multicolumn{5}{c|}{\textbf{Evaluation Metrics (Relative to Full Model)}} & \multirow{2}{*}{\textbf{Avg. Score}} \\
    & \textbf{Cog. Loop} & \textbf{Native MM} & \textbf{Organization} & \textbf{Depth} & \textbf{Relevance} & \textbf{Alignment} & \textbf{Synergy} & \\ \midrule
  GPT-4.1 (W/Search) & $\times$ & $\times$ & 0.4722 & 0.4080 & 0.4875 & 0.3519 & 0.3400 & 0.4119 \\
  CogGen-no-review & $\times$ & \checkmark & 0.4611 & 0.4548 & 0.4889 & \textbf{0.5002} & 0.4356 & 0.4681 \\
  CogGen-TwoStage & \checkmark & $\times$ & 0.4893 & \textbf{0.5167} & 0.4944 & 0.4627 & 0.4890 & 0.4904 \\
   \midrule
  \textbf{CogGen} & \checkmark & \checkmark & \textbf{0.4986} & 0.5000 & \textbf{0.4986} & 0.5000 & \textbf{0.5000} & \textbf{0.4994} \\ \bottomrule
  \end{tabular}
  \caption{\textbf{Ablation Study Results} on OWID dataset. \textbf{Cog. Loop,} Cognitive Loop denotes the reviewer-driven dynamic modification, and \textbf{Native MM,} Native Multimodality refers to the synchronous text-image collaborative planning (via AVR). Scores denote Relative Advantage using CogGen as the reference.}
\label{tab:ablation_final}
\end{table*}
  
\section{Results and Analysis}
\label{sec:results}

\subsection{Main Experimental Results}
\label{sec:main_results}

Table \ref{tab:main_results} presents the Relative Advantage Scores calculated based on the CLEF evaluation metrics. The experimental results show that CogGen exhibits significant advantages in tests on both the OWID and WildSeek datasets.

On the OWID dataset, CogGen demonstrates strong generation capabilities, achieving evaluation scores approaching the Human Gold-Standard while significantly outperforming baseline models such as Multimodal Deep Researcher and WriteHere. Regarding multimodal alignment, although CogGen slightly trails human experts, it secures superior synergy scores compared to all baselines. This advantage is driven by the AVR strategy, which enables iterative coordination between textual and visual planning. Notably, CogGen surpasses human references in Depth. 
We attribute this to that CogGen explicitly provides broader causal context and background information, resulting in higher informational density.

Experiments on the WildSeek dataset further verify the generalization ability of CogGen. With Gemini Deep Research as the reference benchmark, CogGen achieves the highest scores in all five evaluation dimensions. Although Gemini reports narrow the score gap in the multimodal dimension through rich tabular content, their shortcoming of lacking adaptive narrative ability is still obvious. In contrast, baseline models such as WriteHere adopt a recursive decomposition strategy but lack a retroactive rewriting mechanism, leading to fragmented report structures. In comparison, CogGen relies on a hierarchical recursive mechanism to dynamically adjust the outline, ultimately achieving comprehensive leadership in all five dimensions.

\subsection{Ablation Study}
\label{sec:analysis}
Table \ref{tab:ablation_final} details the comparative performance of CogGen against a Retrieval-Augmented GPT-4.1 baseline and its own ablation variants. In direct comparison, the full CogGen framework demonstrates a comprehensive advantage over the GPT-4.1 baseline across all evaluation metrics. Most notably, we observe significant gains in Depth and Synergy, validating that our recursive architecture outperforms standard linear RAG workflows in handling complex, multimodal synthesis tasks.

To isolate the specific contributions of our architectural innovations, we conducted ablation studies focusing on two critical mechanisms: (1) Cognitive Loop: reviewer-driven recursive modification. (2) Native Multimodality: text-image collaborative planning via the AVR strategy. We implemented two variants to verify whether these mechanisms are essential for enhancing content quality and ensuring high-quality visual integration.

\textbf{CogGen-no-review}: This variant removes the recursive modification mechanism for the outline, retaining only the iterative retrieval and parallel section writing functions. Experimental results indicate that after removing the recursive modification mechanism, the model's scores in the three metrics of Organization, Depth, Synergy all show a significant decline; while the scores of Alignment and Relevance remain basically stable. This result shows that the core role of the review module is to improve the global content organization ability and analysis performance of the report, while the writing quality of local content mainly depends on the inherent capabilities of the model.

\textbf{CogGen-TwoStage}: This variant removes the AVR-based image-text coordination from the planning and generation phases. It employs a 'text-first, image-later' strategy, where the model first generates a plain text report before embedding AVR-driven visualizations for final rendering. Experimental data shows that this two-stage generation pipeline results in the most significant drop in the Alignment metric, because the post-inserted images struggle to achieve coherent semantic alignment with the textual content. Synergy has a slight decline, as the text-derived visualizations still effectively reduce cognitive load despite lacking explicit alignment. Notably, the Content Depth of this two-stage variant even surpasses that of the full model. This result aligns with our hypothesis: decoupling visual constraints reduces the cognitive load during text generation, enhancing the depth of analysis.

\subsection{Human Evaluation}
\label{sec:human_eval}
We further conducted a blinded head-to-head human evaluation of CogGen against the baseline model Multimodal DeepResearcher (MMDR) and the proprietary closed-source model Gemini Deep Research on the WildSeek dataset, with assessments carried out across four dimensions: Depth, Alignment, Synergy, and Overall Quality.

CogGen achieved a dominant 90\% win rate over Multimodal DeepResearcher in terms of Overall Quality. Notably, against Gemini Deep Research, CogGen maintained a significant edge in both Overall Quality (75\% win rate) and Multimodal Synergy (80\% win rate); additionally, despite being built on a weaker base model, CogGen attained comparable reasoning depth to Gemini (50\% win rate). Human evaluation results and automatic evaluation results in Table~\ref{tab:main_results} consistently validate the effectiveness of the proposed hierarchical recursive framework CogGen (see Appendix \ref{app:human_h2h} for details). Bootstrap significance analysis ($B{=}10{,}000$) further confirms that CogGen is the only system with no significant difference from the human reference level ($p{=}0.88$; Appendix~\ref{app:bootstrap}).
Additionally, factuality evaluations confirm CogGen's reliability, achieving the highest citation precision and human-verified supported rate among all compared systems (Appendix~\ref{app:factuality}).

\subsection{Efficacy of AVR}
\label{sec:avr_efficacy}

To validate the Abstract Visual Representation (AVR), we compare it with the Formal Description of Visualization (FDV) used in MMDR. By capturing only semantic intent rather than full visual specification, AVR significantly reduces the cognitive burden on the Writer, freeing it from visual design duties---a factor we argue mitigates the Dual-Task Interference reflected in MMDR's lower scores across all dimensions in Table~\ref{tab:main_results}. The ablation results in Section~\ref{sec:analysis} corroborate this hypothesis.

Beyond reducing cognitive load, AVR's decoupled architecture directly addresses the critical issue of chart data hallucination. As shown in Table~\ref{tab:hallucination}, while AVR without verification exhibits hallucination rates comparable to FDV (67\% vs. 60\%), its lightweight format provides a natural insertion point for a Post-Rendering Audit. By cross-checking the rendered data points against the knowledge base, this verification-in-the-loop mechanism substantially reduces the final hallucination rate to 28\%. This demonstrates that AVR is a structural enabler for reliable multimodal generation. For detailed token-level cognitive load analysis, see Appendix~\ref{sec:design_philosophy}.

\begin{table}[t]
\centering
\small
\renewcommand{\arraystretch}{1.2}
\setlength{\tabcolsep}{4pt}
\begin{tabular}{l|cc}
\toprule
\textbf{Configuration} & \textbf{Halluc.} & \textbf{No Halluc.} \\
\midrule
FDV (MMDR) & 60\% & 40\% \\
AVR w/o verification & 67\% & 33\% \\
\textbf{AVR + verification} & \textbf{28\%} & \textbf{72\%} \\
\bottomrule
\end{tabular}
\caption{Chart data hallucination rates across visualization strategies. AVR without verification has comparable hallucination rates to FDV, but the decoupled architecture enables a Post-Rendering Audit that substantially reduces hallucination.}
\label{tab:hallucination}
\end{table}

\section{Conclusion}
This paper presents CogGen, a cognitively inspired framework that overcomes the linear execution constraints of current deep research agents. By integrating a Hierarchical Recursive Architecture with a Parameterized Placeholder Mechanism, CogGen enables non-linear logic restructuring and synergistic multimodal integration. Our evaluation via the CLEF framework and OWID benchmark demonstrates that CogGen achieves performance comparable to human experts in analytical depth and multimodal synergy. These findings validate the efficacy of cognitive architectures in evolving LLMs from linear executors into autonomous, recursive researchers.

\section*{Acknowledgments}
We thank the anonymous reviewers and the area chair for their constructive feedback, which significantly improved this paper. This work is supported by the NSFC (No. 62376120, 62576163).

\section*{Limitations}
While CogGen introduces parallelized generation to improve efficiency, the introduced recursive mechanisms incur additional computational overhead. Furthermore, constrained by current generation and rendering bottlenecks, there remains a quality gap between our automated charts and those curated by human experts. Additionally, the current rendering pipeline deliberately restricts the Render Agent to high-level declarative libraries (ECharts and Mermaid) to ensure stability; this design choice limits the expressiveness for highly customized scientific visualizations achievable through imperative programming. 

\section*{Ethical considerations}
We prioritize ethical responsibility throughout the framework's development. Regarding information veracity, we acknowledge that despite verification mechanisms, LLMs may produce hallucinations; thus, generated reports should serve as references requiring human oversight rather than absolute truths, and we caution against potential misuse for disinformation. In terms of data privacy, we rigorously filtered our dataset to exclude Personally Identifiable Information (PII)  and utilized commercial APIs in compliance with usage policies. Finally, our human evaluation involved graduate student volunteers who participated with full knowledge of the study's purpose and without financial compensation, ensuring adherence to ethical standards for user studies.


\bibliography{agent}

\clearpage

\appendix
\section{Theoretical Analysis of Convergence}
\label{sec:Appendix_Convergence}

In this section, we provide a formal analysis of the convergence properties of CogGen's parallel-recursive architecture. We model the report generation process as a discrete dynamical system and analyze how the proposed \textit{Reviewer Gating Mechanism} acts as a monotonic filter, promoting convergence toward a stable local optimum. Under the premise of noisy LLM judgments, this mechanism is best understood as an empirically effective heuristic rather than a strict theoretical guarantee.

\subsection{System Modeling}

Let $\mathcal{S}$ be the state space of all possible report drafts. A state $S_t \in \mathcal{S}$ at iteration $t$ is defined by the tuple $(\mathcal{O}^{(t)}, \mathcal{C}^{(t)})$, representing the current outline and content.
We define an \textit{Inconsistency Energy Function} $E: \mathcal{S} \rightarrow \mathbb{R}_{\geq 0}$, which quantifies the total logical conflict and quality deficit within a report.
\begin{equation}
    E(S_t) = \sum_{i=1}^{N} \text{Loss}_{\text{local}}(c_i) + \lambda \sum_{i,j} \text{Conflict}(c_i, c_j)
\end{equation}
where $\text{Loss}_{\text{local}}$ quantifies the quality deficit of a single section, and $\text{Conflict}$ represents logical contradictions between sections $i$ and $j$.
A perfect report corresponds to a state $S^*$ where $E(S^*) \to 0$.

\subsection{Convergence of Deferred Resolution}

The core challenge in recursive writing is \textit{Contextual Oscillation}, where a local repair in section $i$ increases the conflict with section $j$, causing $E(S_{t+1}) > E(S_t)$ and leading to limit cycles (infinite loops).
CogGen addresses this via the \textbf{Deferred Resolution Strategy} and \textbf{Global Review Gating}.

\textbf{Proposition 1 (Convergence under Idealized Gating).}
\textit{The CogGen generation process converges to a local optimum if the Reviewer Agent $A_r$ enforces a strict energy descent condition.}

\textit{Proof Sketch.}
In the parallel phase, the Writer generates a candidate set of updates $\Delta S$. The Reviewer $A_r$ does not accept these updates individually. Instead, it evaluates the aggregated next state $S'_{t+1}$.
The Gating Mechanism (Eq.~\ref{eq:gating}) accepts the transition $S_t \to S_{t+1}$ if and only if:
\begin{equation}
\label{eq:gating}
    Q(S'_{t+1}) - Q(S_t) \geq \epsilon
\end{equation}
where $Q$ is the quality score estimated by the LLM (an inverse proxy for Energy $E$) and $\epsilon > 0$ is a minimum improvement threshold.
Since the state space of meaningful reports is finite and bounded, and the quality score $Q$ is bounded from above (e.g., by the maximum context window capacity or logical completeness), a strictly increasing sequence $Q(S_0), Q(S_1), \dots$ must converge to a fixed point where no further improvement $\geq \epsilon$ is possible.
At this point, the system terminates.

\subsection{Complexity Advantage}

Unlike serial backtracking, which suffers from worst-case exponential complexity due to cascading edits ($O(k^N)$ in naive recursive repair), CogGen's parallel update dampens the complexity.
By calculating updates for all defect nodes simultaneously, CogGen approximates the gradient descent direction of the Energy function $E$ over the entire report structure.
Assuming the decoupling of sections allows for independent convergence rates, the time complexity is dominated by the slowest converging section rather than the sum of all revisions:
\begin{equation}
    T_{\text{CogGen}} \approx \max_{i} (m_i) \cdot T_{\text{step}}
\end{equation}
where $m_i$ is the number of revisions for section $i$. This represents a significant speedup over the serial cumulative time $\sum m_i \cdot T_{\text{step}}$.

\textbf{Empirical Validation.}
These theoretical convergence properties are corroborated by the execution statistics presented in Appendix~\ref{app:data_analysis}. Specifically, the low \textit{Global Restructure Rate} (16.0\%) and the rapid generation latency (3.61 min) detailed in Table~\ref{tab:internal_stats} validate that the parallel architecture effectively suppresses worst-case oscillation, aligning with our complexity analysis.

\section{Experimental Analysis}
\label{app:data_analysis}

In this section, we analyze the computational efficiency of CogGen. We first provide a formal specification of the parallel execution mechanism, then benchmark the generation latency against baseline models (Table~\ref{tab:model_comparison}), and finally provide a granular decomposition of CogGen's internal execution to explain the source of latency and validate the system's architectural stability (Table~\ref{tab:internal_stats}).

\subsection{Formal Specification of Parallel Execution}
\label{sec:parallel_formal}

This subsection provides the formal specification of CogGen's parallel micro-cycle execution, including the write isolation constraints and knowledge base synchronization protocol referenced in Section~\ref{sec:Micro_Mechanism}.

\paragraph{Write Isolation Constraint.}
Each parallel thread $\text{Thread}_s$ operates as a \textit{read-only observer} of the global outline $\mathcal{O}^{(t)}$ and all other sections' content $C_{j \neq s}^{(t)}$. No thread may modify the outline or any other section's content during execution. This invariant is enforced architecturally: threads receive a frozen copy of $\mathcal{O}^{(t)}$ at the start of each macro-iteration, eliminating race conditions by construction.

\paragraph{Hierarchical Knowledge Base Protocol.}
The knowledge base $K$ is partitioned into a \textbf{Global Tier} $K_g$ and a \textbf{Local Tier} $K_s$. The global tier is a shared, immutable snapshot constructed during macro planning; all threads read from the same $K_g$. The local tier is a thread-local cache where $\text{Thread}_s$ stores evidence retrieved during its micro-cycle retrieval phase, invisible to other threads. The effective knowledge available to $\text{Thread}_s$ is therefore $K_{\text{eff}}(s) = K_g \cup K_s$, where $K_s \cap K_{s'} = \emptyset$ for $s \neq s'$. This isolation prevents irrelevant noise from propagating between unrelated chapters.

\paragraph{Execution Sequence.}
The parallel micro-cycle proceeds through three phases. First, in the \textit{Dispatch and Parallel Planning} phase, the macro controller broadcasts $\mathcal{O}^{(t)}$ and $K_g$ to all threads. Each $\text{Thread}_s$ independently performs targeted retrieval and generates a section-level plan for $o_s \in \mathcal{O}^{(t)}$, populating its local cache $K_s$. Second, a synchronous \textit{Coarse-Grained Plan Aggregation} step consolidates all section-level plans, performing cross-section deduplication and boundary adjustment to eliminate redundancy \textit{before} writing begins. This lightweight, structure-level consistency pass ensures that parallel plans do not overlap or conflict at the outline level. Third, in the \textit{Parallel Writing} phase, each $\text{Thread}_s$ composes the content $c_s$ based on its consolidated plan, executing the recursive Write--Review micro-loop. A barrier synchronization ensures all threads complete before the unified draft $\mathcal{C}^{(t)} = \{c_s \mid \forall s\}$ is assembled.

\paragraph{Two-Tier Consistency Architecture.}
Once the complete draft is available, the Reviewer $A_r$ performs a \textit{Fine-Grained Global Review}---a holistic, content-level evaluation that detects cross-section logical conflicts, factual inconsistencies, and structural imbalances that the coarse-grained plan aggregation cannot capture---and produces the feedback signal $\Delta^{(t)}$. The transition $\mathcal{O}^{(t)} \to \mathcal{O}^{(t+1)}$ is accepted only if the quality improvement exceeds the threshold $\epsilon$ (Eq.~\ref{eq:gating}). This two-tier design---coarse-grained aggregation \textit{before} writing and fine-grained review \textit{after} writing---ensures that no partial state is ever observed by the Reviewer, enabling deterministic conflict resolution while minimizing redundant generation effort.

\subsection{Latency Analysis}
\label{sec:latency_comp}

Table~\ref{tab:model_comparison} compares the average generation time across five report generation frameworks. We observe a distinct stratification in temporal performance, which correlates with the depth of information processing and the retrieval strategies employed.

\begin{table}[t]
    \centering
    \footnotesize
    \renewcommand{\arraystretch}{1.2}
    \setlength{\tabcolsep}{6pt} 
    
    \begin{tabular}{lr} 
        \toprule
        \textbf{Model} & \textbf{Time (min)} \\
        \midrule
        \multicolumn{2}{l}{\textit{Linear Models}} \\ 
        STORM & 1.54 \\
        CO-STORM & 3.55 \\
        \midrule
        \multicolumn{2}{l}{\textit{Recursive Models}} \\ 
        Multimodal DeepResearcher & 10.46 \\
        WriteHere & 14.75 \\
        \textbf{CogGen (Ours)} & \textbf{20.50} \\
        \bottomrule
    \end{tabular}
    \caption{Efficiency comparison on the OWID dataset ($N=40$).}
    \label{tab:model_comparison}
\end{table}

\begin{table}[t]
    \centering
    \small
    \renewcommand{\arraystretch}{1.2}
    \setlength{\tabcolsep}{12pt}
    \begin{tabular}{lr}
        \toprule
        \textbf{Metric} & \textbf{Value} \\
        \midrule
        \multicolumn{2}{l}{\textit{Time \& Latency}} \\
        Retrieval Duration & 16.89 min (82.4\%) \\
        Generation Duration & \textbf{3.61 min} (17.6\%) \\
        Retrieval Latency / req & 78.05 s \\
        Generation Latency / req & 5.48 s \\
        \midrule
        \multicolumn{2}{l}{\textit{Resource Allocation}} \\
        Avg. Cost & $\approx \$4.80$ \\
        Total Tokens & 5.01 M \\
        \quad \textit{- Retrieval Phase} & $\approx 80\%$ \\
        \quad \textit{- Generation Phase} & $\approx 20\%$ \\
        \midrule
        \multicolumn{2}{l}{\textit{Execution Dynamics}} \\
        Plan Modifications & \textbf{2.39} \\
        Content Modifications & \textbf{0.43} \\
        Zero-Shot Success & 71.1\% \\
        Restructure Rate & 16.0\% \\
        \bottomrule
    \end{tabular}
    \caption{Internal execution statistics of CogGen. Data represents averages from the OWID dataset ($N=40$).}
    \label{tab:internal_stats}
\end{table}

\paragraph{Retrieval Pipelines and Fidelity.}
While all frameworks in our evaluation utilize the Tavily Search as the unified retrieval source, their post-retrieval processing strategies diverge significantly to align with their respective architectural goals.

\textbf{Snippet-based Processing.} Baselines such as STORM and WriteHere are designed to optimize for response speed. They typically ingest search snippets or RAG-retrieved chunks directly. While efficient, we argue that for long-form report generation, relying solely on snippets carries the risk of \textit{contextual fragmentation}, where disconnected text segments may induce logical inconsistencies or hallucinations during synthesis.
    
\textbf{Full-Content Summarization.} In contrast, CogGen explicitly implements a \textit{Crawler-Summarizer Pipeline} (reading full web pages and summarizing via LLM), aligning with the technical framework of deep research agents like Tongyi DeepResearch \cite{tongyideepresearchteam2025tongyideepresearchtechnicalreport}. We treat this computationally intensive step as a necessary ``Denoising and Verification'' layer. By digesting the complete document context before synthesis, the model filters out irrelevant noise and ensures better logical coherence, effectively mitigating the hallucination risks inherent in snippet-stitching approaches.

\textbf{Impact on Quality Assessment.} 
Crucially, this comprehensive ingestion strategy does not artificially inflate the structural or multimodal evaluation metrics (e.g., Organization, Alignment) used in CLEF. Instead, its primary function is to \textbf{mitigate hallucinations}. By ensuring that the model reasons over verified summaries rather than fragmented snippets, we guarantee that the high scores achieved in the ``Depth'' dimension (Table~\ref{tab:main_results}) reflect genuine analytical capability rather than plausible-sounding fabrications. This ensures a rigorous and valid quality comparison where CogGen's advantage stems from its recursive architecture, not just data quantity.

\paragraph{Latency Attribution and Architectural Speed.}
Table~\ref{tab:model_comparison} indicates that CogGen's total latency (20.50 min) is higher than the snippet-based baselines. It is crucial to note that 82.4\% of this time is allocated to the heavy Ingestion Phase (full-page reading and summarization), a deliberate design choice to prioritize information fidelity over raw speed.

Most importantly, when isolating the Reasoning and Generation Phase (Table~\ref{tab:internal_stats}), CogGen completes the complex multimodal planning and writing in only 3.61 minutes. This confirms that our Gated Parallelism mechanism effectively solves the bottleneck of recursive generation, achieving a throughput significantly higher than serial recursive baselines like WriteHere (14.75 min).
Future deployments could mitigate retrieval latency by employing specialized lightweight summarization models instead of general-purpose LLMs.

\begin{table*}[t]
\centering
\small
\setlength{\tabcolsep}{5pt}
\begin{tabular}{l|ccccc|c}
\toprule
\textbf{Model} & \textbf{Organization} & \textbf{Depth} & \textbf{Relevance} & \textbf{Alignment} & \textbf{Synergy} & \textbf{Avg. Score} \\ \midrule
\multicolumn{7}{c}{\textbf{\textit{Evaluator I: Doubao-Seed-1.6 (Judge)}}} \\ \midrule
Gemini Deep Research (Ref) & 0.5000 & 0.5000 & 0.5000 & 0.5150 & 0.5000 & 0.5030 \\
Multimodal DeepResearcher & 0.3938 & 0.3536 & 0.3641 & 0.2300 & 0.2433 & 0.3170 \\
WriteHere & \underline{0.5466} & \underline{0.5382} & \underline{0.5261} & \underline{0.5352} & \underline{0.5005} & \underline{0.5293} \\ \midrule
\textbf{CogGen (Ours)} & \textbf{0.5591} & \textbf{0.5528} & \textbf{0.5500} & \textbf{0.6548} & \textbf{0.6762} & \textbf{0.5986} \\ \midrule
\multicolumn{7}{c}{\textbf{\textit{Evaluator II: Claude-Sonnet-4 (Judge)}}} \\ \midrule
Gemini Deep Research (Ref) & 0.5028 & 0.5028 & 0.5000 & \underline{0.5833} & \underline{0.6494} & 0.5477 \\
Multimodal DeepResearcher & 0.3080 & 0.2991 & 0.3080 & 0.2017 & 0.1967 & 0.2627 \\
WriteHere & \textbf{0.5610} & \underline{0.5609} & \textbf{0.5473} & 0.5298 & 0.5562 & \underline{0.5510} \\ \midrule
\textbf{CogGen (Ours)} & \underline{0.5474} & \textbf{0.5715} & \underline{0.5334} & \textbf{0.7181} & \textbf{0.6572} & \textbf{0.6055} \\ \bottomrule
\end{tabular}
\caption{\textbf{Robustness Analysis on WildSeek Dataset across Evaluators.} Comparison of model performance when evaluated by different judge models: Doubao-Seed-1.6 (top) and Claude-Sonnet-4 (bottom). \textbf{Bold} highlights the best result, and \underline{underlined} marks the second best.}
\label{tab:appendix_evaluator_comparison}
\end{table*}

\subsection{Internal Dynamics and Stability}
\label{sec:internal_stats}

To validate the Deferred Update mechanism proposed in Section~\ref{sec:Micro_Mechanism}, we analyze the internal behavioral statistics of CogGen. Table~\ref{tab:internal_stats} details the resource consumption and modification patterns.

\paragraph{Planning Flux vs. Writing Stability.}
The statistics reveal a functional decoupling between planning and writing. The Planner exhibits high activity (2.39 revisions/section), absorbing the uncertainty of the task. In contrast, the Writer demonstrates high stability (0.43 revisions/section) with a 71.1\% zero-shot success rate. The 5.6:1 ratio between plan and content revisions provides empirical evidence that the hierarchical architecture effectively transforms a complex reasoning problem into a deterministic execution task. 

Crucially, this stability does not imply rigidity. The Global Restructure rate (16.0\%) indicates that while the local writing link prioritizes efficiency optimization, the global planning link maintains the flexibility to adapt to logical conflicts discovered during execution. This hierarchical dynamism ensures that the system avoids the ``tunnel vision'' typical of linear models while minimizing the latency cost of full recursion.

\subsection{Backward Restructuring Analysis}
\label{app:backward_restructuring}

To provide concrete evidence of the backward restructuring mechanism described in Section~\ref{sec:Macro_Mechanism}, we analyze restructuring events observed across the evaluated reports.

\paragraph{Frequency and Outcomes.}
Across all evaluated reports, 13.3\% of outline modifications involve backward restructuring---cases where downstream content discoveries trigger retroactive changes to the global outline. We manually examined all observed backward restructuring events and found no harmful updates. All cases involved structural optimizations such as eliminating cross-section redundancy and adjusting section boundaries, with consistent Reviewer decision direction.

\paragraph{Representative Example.}
In a report on ``What are the safest and cleanest sources of energy?'', the Reviewer identified content overlap between \S2.1's comprehensive ranking and Chapter~6's summary synthesis during the macro-cycle review, triggering backward restructuring. Table~\ref{tab:backward_example} presents the original outline and the Planner's targeted modification instructions.

\begin{table*}[t]
\centering
\small
\renewcommand{\arraystretch}{1.3}
\begin{tabularx}{\textwidth}{@{} l X @{}}
\toprule
\textbf{Stage} & \textbf{Content} \\
\midrule
\textit{Before} &
\texttt{2.1 Comparative Ranking of Lifecycle GHG Emissions} --- Analyze lifecycle GHG emissions for major energy sources using quantified values for coal, oil, gas, nuclear, wind, solar\ldots \\
\midrule
\textit{After} &
\textbullet\ Trim paragraphs that broadly summarize which sources are ``dirtiest'' or ``cleanest''---leave detailed ranking and synthesis for \S6. \newline
\textbullet\ Keep detailed lifecycle GHG data, methodology, and regional/technological variability analysis. \newline
\textbullet\ Remove summary statements duplicating \S6's synthesis. \\
\bottomrule
\end{tabularx}
\caption{Backward restructuring example: the Planner's revision of \S2.1 to eliminate cross-section redundancy with Chapter~6.}
\label{tab:backward_example}
\end{table*}

After modification, Section~2.1 retained detailed lifecycle emission data and methodological analysis, while comprehensive conclusions were deferred to the final chapter, eliminating cross-section redundancy.

\section{Detailed Evaluation}
\label{app:detailed_eval}

\subsection{Evaluation Across Different Models}
In the main text, we adopt GPT-5 as the primary evaluation judge owing to its superior reasoning capabilities and strong alignment with human preferences. To mitigate potential biases induced by the choice of a single evaluation model and to verify the cross-model robustness of our results, we further conducted experiments on the WildSeek dataset using two distinct state-of-the-art LLMs as alternative judges: \textbf{Doubao-Seed-1.6} and \textbf{Claude-Sonnet-4}. The comparative evaluation results under the CLEF framework are presented in Table~\ref{tab:appendix_evaluator_comparison}. It is important to note that the model employed in our generation process (CogGen) is completely independent of these judge models, ensuring a blind evaluation setting.

As shown in Table~\ref{tab:appendix_evaluator_comparison}, while the absolute scoring distributions vary between judges (e.g., Claude-Sonnet-4 tends to assign higher baseline scores to the reference), the relative performance trends remain highly consistent. \textbf{CogGen} maintains the highest Overall Average Score across all evaluators. Notably, in critical multimodal metrics such as \textit{Alignment} and \textit{Synergy}, CogGen consistently outperforms baselines by a significant margin regardless of the evaluator used. These results confirm that our method's superiority is intrinsic to the generated content quality and is robust to the variations in evaluation models.

\begin{figure*}[h!] 
    \centering
    \includegraphics[width=\textwidth]{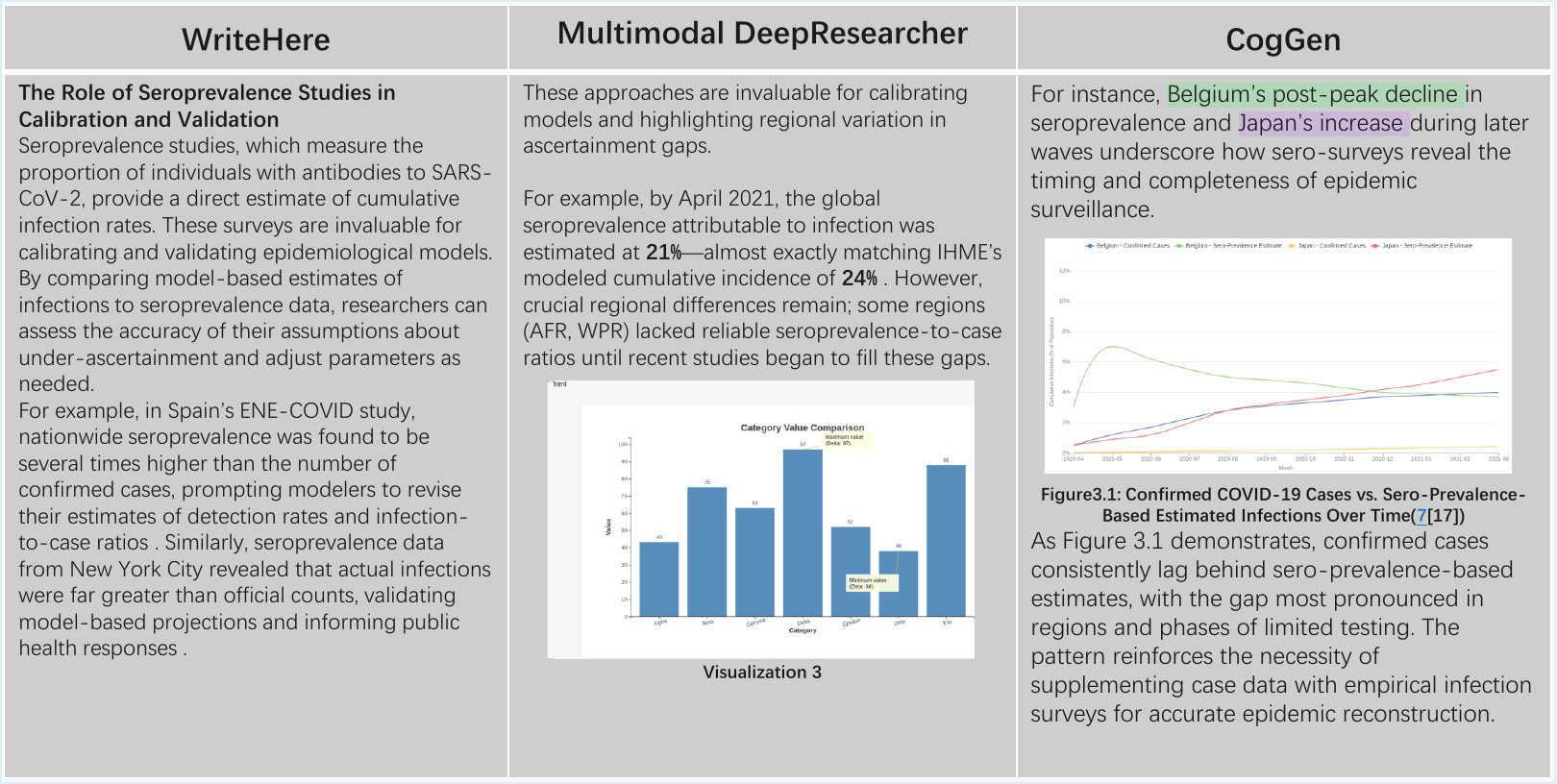} 
    \caption{Qualitative Comparison of Cross-Modal Alignment Performance:
    The left panel displays the output of the baseline model WriteHere; the middle panel presents the generated results of Multimodal DeepResearcher; and the right panel shows the output of our proposed CogGen method. We adopt a color-coded highlighting approach to mark the correspondences between textual content and visual elements.}
    \label{fig:appendix_comparison}

\end{figure*}

\subsection{Human Comparative Evaluation}
\label{app:human_h2h}

To rigorously validate CogGen's effectiveness, we conducted a blinded head-to-head human evaluation on WildSeek, comparing against two baselines: (1) Multimodal DeepResearcher (MMDR), a multimodal baseline  using a linear workflow; and (2) Gemini Deep Research, a proprietary commercial system, to benchmark overall performance.

\subsubsection{Setup}
\textbf{Evaluation Protocol.} We evaluated all 20 WildSeek queries without sampling to eliminate selection bias. A blinded annotator assessed each report pair across four dimensions: Overall Quality, Alignment, Synergy, and Depth. Statistical significance was assessed using the Wilcoxon signed-rank test (ties excluded).

\subsubsection{Results}
Tables~\ref{tab:human_vs_mmdr} and~\ref{tab:human_vs_gemini} present the comparative results.

\begin{table}[t]
\centering
\small
\begin{tabular}{l|cc}
\toprule
\textbf{Dimension} & \textbf{W/T/L} & \textbf{Win\%} \\
\midrule
Overall Quality & 18/1/1 & \textbf{90.0}$^{**}$ \\
Content Depth & 19/0/1 & \textbf{95.0}$^{**}$ \\
Visual-Text Alignment & 16/2/2 & \textbf{80.0}$^{**}$ \\
Multimodal Synergy & 16/2/2 & \textbf{80.0}$^{**}$ \\
\bottomrule
\end{tabular}
\caption{Human evaluation: CogGen vs. MMDR ($N{=}20$). W/T/L: Win/Tie/Loss. $^{**}$: $p{<}0.01$.}
\label{tab:human_vs_mmdr}
\end{table}

\begin{table}[t]
\centering
\small
\begin{tabular}{l|cc}
\toprule
\textbf{Dimension} & \textbf{W/T/L} & \textbf{Win\%} \\
\midrule
Overall Quality & 15/1/4 & \textbf{75.0}$^{*}$ \\
Visual-Text Alignment & 16/0/4 & \textbf{80.0}$^{**}$ \\
Multimodal Synergy & 16/1/3 & \textbf{80.0}$^{**}$ \\
Content Depth & 10/3/7 & 50.0 \\
\bottomrule
\end{tabular}
\caption{Human evaluation: CogGen vs. Gemini ($N{=}20$). $^{*}$: $p{<}0.05$; $^{**}$: $p{<}0.01$.}
\label{tab:human_vs_gemini}
\end{table}

\textbf{AVR Mechanism Validation.}
Table~\ref{tab:human_vs_mmdr} demonstrates CogGen's substantial advantage over MMDR across all dimensions (win rates $\geq$80\%). The 95\% win rate in Depth validates our multimodal reasoning framework, while consistent 80\% wins in alignment and synergy dimensions empirically confirm AVR's effectiveness in bridging the reasoning-rendering semantic gap compared to MMDR's implementation.

\textbf{Gemini Comparison.}
Compared with the Gemini Deep Research (Gemini) (see Table~\ref{tab:human_vs_gemini}), CogGen achieves a statistically significant advantage in both Overall Quality (75\% win rate, \(p<0.05\)) and Multimodal Dimension (80\% win rate, \(p<0.01\)). We draw two core findings: (1)Multimodal Advantage: CogGen's AVR mechanism enables precise, context-aware chart placement; while Gemini generates abundant tables, they often lack contextual relevance. (2)Reasoning Parity: CogGen ties with Gemini (50\% win rate) in the Content Depth dimension. This demonstrates that the hierarchical recursive framework proposed in our study not only excels in multimodal fusion performance, but also matches the reasoning capabilities of proprietary commercial systems.

\begin{table*}[h]
\centering
\small
\begin{tabularx}{\textwidth}{@{} l X @{}}
\toprule
\textbf{Dimension \& CTML Principles} & \textbf{Evaluation Focus \& Cognitive Goal} \\
\midrule
\multicolumn{2}{l}{\textit{\textbf{Control Dimensions: General Quality}}} \\
\addlinespace[0.2em]
D1: Information Organization & \textbf{Hierarchical structure.} Evaluates if headings and layout guide attention effectively via signaling. \\
\textit{(Signaling, Segmenting)} & (Extraneous Load $\downarrow$) \\
\addlinespace[0.6em]

D2: Content Depth and Insight & \textbf{Causal explanations.} Assesses whether content promotes deep reasoning and schema construction over fact stacking. \\
\textit{(Concreteness, Schema Construction)} & (Germane Load $\uparrow$) \\
\addlinespace[0.6em]

D3: Content Relevance and Adaptation & \textbf{Appropriate complexity.} Ensures content excludes distracting details and matches user intent. \\
\textit{(Coherence, Pre-training, Personalization)} & (Extraneous $\downarrow$, Intrinsic Managed) \\
\midrule
\multicolumn{2}{l}{\textit{\textbf{Core Dimensions: Multimodal Integration}}} \\
\addlinespace[0.2em]
D4: Visual-Text Alignment & \textbf{Tight spatial/semantic integration.} Assesses whether elements are physically and semantically close to reduce split-attention effects. \\
\textit{(Spatial Contiguity)} & (Extraneous Load $\downarrow$) \\
\addlinespace[0.6em]

D5: Multimodal Synergy & \textbf{Complementarity beyond text.} Checks if visuals provide unique information gain (e.g., trends) rather than decoration. \\
\textit{(Multimedia, Redundancy, Image)} & (Extraneous $\downarrow$, Germane $\uparrow$) \\
\bottomrule
\end{tabularx}
\caption{Detailed definitions of CLEF dimensions, mapped to CTML Principles and cognitive load targets.}
\label{tab:clef_dimensions}
\end{table*}

\subsection{Case Study}
\label{sec:appendix_case_study}
Due to space constraints in the main text, we place the qualitative case comparison in the appendix, as illustrated in Figure~\ref{fig:appendix_comparison}. We compared three frameworks—WriteHere, Multimodal DeepResearcher, and CogGen—regarding their descriptive performance on seroprevalence-based approaches. Empirical examples demonstrate that WriteHere generates text-only content, with no quantitative results included in its case descriptions. Multimodal DeepResearcher produces content integrating text and graphics; however, the textual component lacks analytical depth, and there is no logical correlation between the images and text, which instead disrupts the normal reading flow. In contrast, CogGen, the method proposed in this study, conducts a cross-sectional data comparison between Japan and Belgium, employs line charts to intuitively visualize the developmental trends, and achieves tight integration of text and graphics along with targeted in-depth analysis.

\subsection{Bootstrap Significance Analysis}
\label{app:bootstrap}

To rigorously assess statistical significance, we conducted Bootstrap analysis ($B{=}10{,}000$) on the CLEF evaluation results. CogGen is the only system whose overall score shows no significant difference from the human reference level ($p{=}0.88$, 95\% CI fully covering 0.5), whereas all baselines fall significantly below ($p{<}0.001$). The advantage is most pronounced on the multimodal dimensions (Alignment and Synergy), where CogGen outperforms the strongest baseline WriteHere by over 0.09 points ($p{<}0.001$).

\subsection{Cross-Domain Evaluation}
\label{app:cross_domain}

To verify that CogGen's advantages are not overfit to the original OWID topic distribution (concentrated in Health \& Medicine at 32.5\% and Economics \& Development at 17.5\%), we collected 10 additional multimodal reports spanning previously underrepresented domains including Democracy/Governance, Social Media/Digital Technology, Immigration/Demographics, Financial Technology, Media/Public Perception, and Gender/Demography. Table~\ref{tab:cross_domain} presents the evaluation results.

\begin{table*}[t]
\centering
\small
\renewcommand{\arraystretch}{1.2}
\setlength{\tabcolsep}{5pt}
\begin{tabular}{l|ccccc|c}
\toprule
\textbf{Model} & \textbf{Organization} & \textbf{Depth} & \textbf{Relevance} & \textbf{Alignment} & \textbf{Synergy} & \textbf{Avg. Score} \\
\midrule
\textbf{CogGen (Ours)} & \textbf{0.500} & \textbf{0.590} & 0.472 & \textbf{0.449} & \textbf{0.421} & \textbf{0.486} \\
Multimodal DeepResearcher & 0.403 & 0.475 & 0.375 & 0.179 & 0.170 & 0.320 \\
WriteHere & 0.489 & 0.547 & \textbf{0.511} & 0.438 & 0.378 & 0.473 \\
\bottomrule
\end{tabular}
\caption{Content quality evaluation on 10 additional cross-domain reports. Scores are CLEF Relative Advantage.}
\label{tab:cross_domain}
\end{table*}

The results are consistent with the main experiment trends: CogGen maintains the overall lead (Avg.\ 0.486), with particularly significant advantages on multimodal dimensions (Alignment and Synergy). This confirms that the hierarchical recursive architecture generalizes across diverse domains.

\section{CLEF: Cognitive Load Evaluation Framework Details}
\label{app:clef}

\subsection{Theoretical Foundation}

CLEF is grounded in two complementary theories:

\paragraph{Cognitive Load Theory (CLT)}
CLT identifies three types of cognitive load: \textit{intrinsic load} (content difficulty), \textit{extraneous load} (presentation burden, to be minimized), and \textit{germane load} (schema construction effort, to be maximized)~\cite{sweller1994cognitive}.

\paragraph{Cognitive Theory of Multimedia Learning (CTML)}
Mayer's CTML operationalizes cognitive principles into measurable design dimensions. CLEF maps these principles to evaluation metrics to assess cognitive burden reduction~\cite{mayer2005cognitive}.

\subsection{Evaluation Dimensions}

We map the evaluation dimensions to specific CTML principles and CLT goals. Table~\ref{tab:clef_dimensions} details the evaluation focus for each dimension.

\subsection{Complete Mapping to CTML Principles}

Table~\ref{tab:clef_dimensions} presents the primary CTML principles associated with each evaluation dimension. To provide a comprehensive view, Table~\ref{tab:ctml_complete_mapping} presents the complete mapping from all 14 CTML principles to CLEF dimensions, clarifying coverage and scope.

\begin{table*}[h]
\centering
\small
\begin{tabularx}{\textwidth}{@{} l l X @{}}
\toprule
\textbf{CTML Principle} & \textbf{CLEF Dimension} & \textbf{Mapping Rationale} \\
\midrule
\multicolumn{3}{l}{\textit{\textbf{Principles Directly Evaluated by CLEF}}} \\
\addlinespace[0.2em]

1. Multimedia Principle & D5 & Assesses whether text-visual combinations provide synergistic information gain beyond text alone. \\
\addlinespace[0.3em]

2. Modality Principle & \textit{N/A} & Concerns audio vs. text; not applicable to static multimodal reports. \\
\addlinespace[0.3em]

3. Redundancy Principle & D5 & Evaluates whether visuals complement text rather than merely repeating it verbatim. \\
\addlinespace[0.3em]

4. Spatial Contiguity & D4 & Measures spatial proximity between related text and visual elements to reduce split-attention. \\
\addlinespace[0.3em]

5. Temporal Contiguity & \textit{N/A} & Concerns synchronization in dynamic media; not applicable to static reports. \\
\addlinespace[0.3em]

6. Coherence Principle & D3 & Checks whether content excludes extraneous, distracting, or irrelevant information. \\
\addlinespace[0.3em]

7. Interactivity Principle & \textit{N/A} & Concerns learner-controlled pacing; not applicable to static report evaluation. \\
\addlinespace[0.3em]

8. Signaling Principle & D1 & Evaluates use of headings, highlighting, and structural cues to guide attention. \\
\addlinespace[0.3em]

9. Segmenting Principle & D1 & Assessed through hierarchical organization and logical content chunking. \\
\addlinespace[0.3em]

10. Pre-training Principle & D3 & Indirectly evaluated via content adaptation to user expertise level. \\
\addlinespace[0.3em]

11. Personalization Principle & D3 & Considered in evaluating whether content tone and complexity match user intent. \\
\addlinespace[0.3em]

12. Concreteness Principle & D2 & Assesses use of examples, analogies, and concrete instantiations in explanations. \\
\addlinespace[0.3em]

13. Voice Principle & \textit{N/A} & Concerns audio narration quality; not applicable to text-based reports. \\
\addlinespace[0.3em]

14. Image Principle & D5 & Evaluates whether images serve functional (not decorative) purposes. \\

\midrule
\multicolumn{3}{l}{\textit{\textbf{Cognitive Load Theory (CLT) Integration}}} \\
\addlinespace[0.2em]

Intrinsic Load & D3 & Managed through appropriate content complexity matching user expertise. \\
\addlinespace[0.3em]

Extraneous Load & D4, D1, D3 & Minimized via spatial integration (D4), clear structure (D1), and coherence (D3). \\
\addlinespace[0.3em]

Germane Load & D5, D2 & Enhanced via meaningful visual integration (D5) and deep explanations (D2). \\

\bottomrule
\end{tabularx}
\caption{Complete mapping from Mayer's 14 CTML principles and 3 CLT load types to CLEF's 5 evaluation dimensions. Principles marked \textit{N/A} are not applicable to static multimodal report evaluation.}
\label{tab:ctml_complete_mapping}
\end{table*}

\paragraph{Coverage Analysis}
CLEF's five dimensions systematically operationalize 11 of the 14 CTML principles. Three principles (Modality, Temporal Contiguity, Voice) are excluded as they specifically address dynamic multimedia (audio/video synchronization) and are not applicable to static text-visual reports. The framework comprehensively addresses all three CLT load types: minimizing extraneous load through D4, D1, and D2; managing intrinsic load via D3; and promoting germane load through D5 and D2.

\subsection{Scoring Mechanism}

\paragraph{Pairwise Comparative Evaluation}
Following best practices~\cite{duDeepResearchBenchComprehensive2025a}, GPT-5 simultaneously evaluates both the model report and a reference report.

\paragraph{Relative Advantage Score}
For each dimension $i$, the relative advantage score is calculated as:
\begin{equation}
R_i = \frac{S_{\text{model}}^{(i)}}{S_{\text{model}}^{(i)} + S_{\text{ref}}^{(i)}} \in [0, 1]
\end{equation}
where $R_i > 0.5$ indicates the model report outperforms the reference. The final score is the average across all dimensions:
\begin{equation}
R_{\text{final}} = \frac{1}{5}\sum_{i=1}^{5} R_i
\end{equation}

\subsection{Implementation}

\paragraph{Prompt Structure}
Prompts are structured to mitigate ``Lost in the Middle'' effects~\cite{liu2024lost}: (1) evaluation rubric (as defined in Table~\ref{tab:clef_dimensions}); (2) interleaved text-image content of both reports; (3) holistic comparative instructions. Images are encoded in base64 to leverage GPT-5's native multimodal capabilities.

\section{Factuality Evaluation}
\label{app:factuality}

To quantify CogGen's factual reliability, we conducted both automated and human-verified evaluations on the WildSeek dataset.

\subsection{Evaluation Methodology}

\paragraph{Automated Citation Evaluation.}
We collected all citations from each system's reports across 20 WildSeek queries (11,291 total citations). For each citation, we crawled the source URL and used an LLM to judge the relevance of the cited content to the corresponding statement, computing \textbf{Citation Precision}.

\paragraph{Human Claim-Level Verification.}
We sampled 5 reports from each system and decomposed the most claim-dense paragraphs into 148 atomic claims. Human annotators independently verified each claim via web search, measuring two metrics: \textbf{Supported Rate} (proportion of claims with supporting web evidence) and \textbf{Citation Accuracy} (proportion of cited sources that actually contain the claimed content).

\subsection{Results}

CogGen achieves the highest scores across all three factuality metrics: Citation Precision of 0.72 (vs.\ WriteHere 0.69, Gemini 0.60), human-verified Supported Rate of 76.3\% (vs.\ 72.7\%, 60.5\%), and Citation Accuracy of 55.3\% (vs.\ 54.5\%, 44.2\%). These results demonstrate competitive factual reliability even without dedicated optimization for this dimension.

\subsection{Ingestion Strategy Ablation}
\label{app:ingestion_ablation}

To disentangle the contributions of retrieval strategy (ingestion) and recursive architecture, we replaced CogGen's full-text summarization strategy with lightweight snippet retrieval. The two configurations differ only in the retrieval stage; the writing model receives context in an identical format.

Switching from full-text summarization to snippet retrieval yields nearly identical CLEF scores (0.4992 vs.\ 0.5019) but sharply reduces the Supported Rate from 76.3\% to 50.0\%, while generation time drops from 20.50 to 6.62 minutes. This reveals a clear separation of concerns: CLEF scores are nearly identical, indicating that CogGen's content quality advantage stems from the hierarchical recursive architecture and AVR mechanism, not the retrieval strategy. However, the Supported Rate drops sharply, confirming that the full-text summarization pipeline is critical for factual accuracy. With 82.4\% of total latency attributable to the retrieval stage---recursive reasoning itself requires only ${\sim}3.6$ minutes---users can flexibly choose between a factuality-first mode (20 min) and a speed-first mode (7 min) depending on the use case.

\section{Visualization Implementation Details}
\label{sec:appendix_rendering}

This appendix provides a comprehensive analysis of the visualization generation module in CogGen, detailing the Abstract Visual Representation (AVR) design, the rendering pipeline, architectural trade-offs compared to related work, and statistical validation on the OWID dataset.

\subsection{AVR-based Decoupled Rendering Pipeline}
\label{sec:avr_pipeline}

As introduced in Table~\ref{tab:viz_placeholder} of the main text, the Abstract Visual Representation (AVR) serves as the intermediate bridge between narrative intent and visual execution. The generation process follows a strict pipeline: the \textbf{Planner} determines the chart intent, the \textbf{Writer} generates the AVR structure, and the \textbf{Render Agent} translates AVR into executable code.

\paragraph{AVR Field Structure.}
To ensure generative stability, the AVR schema is divided into mandatory and optional fields:
\begin{itemize}
    \item \textbf{Fixed Fields (Mandatory):} Required for every visualization to define the core intent. These include \texttt{Title}, \texttt{Chart\_Type}, \texttt{Data\_Source}, and \texttt{Purpose}.
    \item \textbf{Dynamic Fields (Optional):} Context-dependent fields such as \texttt{X\_Axis} and \texttt{Y\_Axis} definitions, which are only generated when the specified \texttt{Chart\_Type} requires coordinate mapping (e.g., Bar Charts) and are omitted for types like Pie Charts or Flowcharts.
\end{itemize}

\paragraph{Rendering Technology Stack.}
While LLMs increasingly demonstrate the ability to generate raw HTML/CSS directly, we deliberately constrain the Render Agent to target specific high-level visualization libraries: \textbf{Mermaid.js} and \textbf{Apache ECharts}. 
\begin{itemize}
    \item \textbf{Implementation Strategy:} Rather than permitting the Render Agent to freely hallucinate HTML structures—which often leads to inconsistent styling and broken layouts—the agent generates configuration code for these libraries.
    \item \textbf{Execution Environment:} The rendering occurs in a browser-based environment. Leveraging established frontend libraries ensures interactive, aesthetically consistent, and functionally robust charts while significantly lowering the coding capability requirement for the LLM.
\end{itemize}

\subsection{Cognitive Load Trade-off and Comparison}
\label{sec:design_philosophy}

Our design philosophy centers on minimizing the Dual-Task Interference for the Writer agent. We explicitly trade off granular control for semantic simplicity.

\paragraph{Comparison with Multimodal DeepResearcher.}
Existing systems like Multimodal DeepResearcher (MMDR) adopt a ``Two-Stage'' rendering strategy using a placeholder known as \textbf{FDV} (Formal Description of Visualization). The FDV is designed to describe every visual detail, including style, color, and layout, with high precision. 
\begin{itemize}
    \item \textbf{The MMDR Limitation:} Our empirical observations indicate that such verbose placeholders impose a substantial cognitive load on the Writer agent. Attempting to perfect visual specifications distracts the model from its primary task of narrative construction, leading to degradation in text quality.
    \item \textbf{The CogGen Advantage:} By offloading styling decisions to the standard themes of ECharts and Mermaid, the AVR allows the Writer to focus solely on \textit{data} and \textit{intent}. This ``lightweight'' representation reduces cognitive overhead, preventing the quality dip observed in MMDR.
\end{itemize}

\paragraph{Quantitative Comparison.}
To quantify the cognitive cost difference, we measured the average token count per visualization placeholder across 50 reports. AVR averages ${\sim}133$ tokens per figure (measured over 339 blocks), while FDV averages ${\sim}773$ tokens (measured over 252 blocks)---a 5.8$\times$ difference. This reduction directly reflects the separation of concerns: AVR answers ``what to show and why'' while delegating ``how to draw'' to the dedicated Render Agent.

\paragraph{Post-Rendering Data Verification Pipeline.}
As discussed in Section~\ref{sec:avr_efficacy}, AVR's decoupled nature enables a Post-Rendering Audit, which is architecturally difficult in FDV's monolithic pipeline. In CogGen, this module operates by parsing the intermediate ECharts JSON generated by the Render Agent and cross-checking the exact coordinate data points against the original source values retrieved in the Knowledge Base $K$. This verification-in-the-loop mechanism is responsible for the significant drop in hallucination rates detailed in Table~\ref{tab:hallucination} of the main text.

\subsection{Statistical Analysis of Generated Visualizations}
\label{sec:viz_stats}

To validate the effectiveness of our multimodal report generation system, we conducted a comprehensive statistical analysis on the visualization outputs from the OWID dataset ($N=40$). Table~\ref{tab:viz_statistics} summarizes the key quantitative metrics.

\begin{table}[t]
    \centering
    \small
    \renewcommand{\arraystretch}{1.2}
    \setlength{\tabcolsep}{10pt}
    \begin{tabular}{lr}
        \toprule
        \textbf{Metric} & \textbf{Value} \\
        \midrule
        \multicolumn{2}{l}{\textit{Generation Performance}} \\
        Total Reports & 40 \\
        Requested Visualizations & 258 \\
        Successfully Generated & 248 \\
        Success Rate & \textbf{96.12\%} \\
        Avg. Visualizations per Report & 6.45 \\
        \midrule
        \multicolumn{2}{l}{\textit{Type Diversity}} \\
        Distinct Chart Types & 22 \\
        Primary Categories & 4 \\
        \bottomrule
    \end{tabular}
    \caption{Visualization generation statistics on the OWID dataset.}
    \label{tab:viz_statistics}
\end{table}

\paragraph{High Generation Reliability.}
The system achieved a 96.12\% success rate across 258 visualization requests, demonstrating robust cross-modal generation capability. Each report contains an average of 6.45 visualizations, indicating that the system effectively integrates visual elements to support textual content. This high reliability validates the architectural design of our multimodal generation pipeline.

\paragraph{Chart Type Distribution.}
Table~\ref{tab:chart_types} presents the distribution of generated chart types across functional categories. The system demonstrates strong diversity, producing 22 distinct chart types spanning statistical analysis, process visualization, geographic mapping, and specialized structural diagrams.

\begin{table}[t]
    \centering
    \footnotesize
    \renewcommand{\arraystretch}{1.15}
    \setlength{\tabcolsep}{5pt}
    \begin{tabular}{lrrr}
        \toprule
        \textbf{Chart Type} & \textbf{Count} & \textbf{\%} & \textbf{Cumulative} \\
        \midrule
        \multicolumn{4}{l}{\textit{Statistical Charts (46.9\%)}} \\
        \quad Bar Chart & 69 & 26.74 & 26.74 \\
        \quad Line Chart & 39 & 15.12 & 41.86 \\
        \quad Area Chart & 13 & 5.04 & 46.90 \\
        \midrule
        \multicolumn{4}{l}{\textit{Relational \& Process (25.6\%)}} \\
        \quad Flowchart & 38 & 14.73 & 61.63 \\
        \quad Heatmap & 7 & 2.71 & 64.34 \\
        \quad Pie Chart & 6 & 2.33 & 66.67 \\
        \quad Timeline & 6 & 2.33 & 68.99 \\
        \quad Scatter Plot & 5 & 1.94 & 70.93 \\
        \quad Sankey & 4 & 1.55 & 72.48 \\
        \midrule
        \multicolumn{4}{l}{\textit{Geographic \& Structural (20.2\%)}} \\
        \quad Map & 18 & 6.98 & 79.46 \\
        \quad Diagram & 17 & 6.59 & 86.05 \\
        \quad Infographic & 10 & 3.88 & 89.92 \\
        \quad Matrix & 8 & 3.10 & 93.02 \\
        \midrule
        \multicolumn{4}{l}{\textit{Specialized (7.0\%)}} \\
        \quad Table & 6 & 2.33 & 95.35 \\
        \quad Roadmap & 4 & 1.55 & 96.90 \\
        \quad Others (6 types) & 8 & 3.10 & 100.00 \\
        \bottomrule
    \end{tabular}
    \caption{Distribution of generated chart types across functional categories.}
    \label{tab:chart_types}
\end{table}

\paragraph{Dominance of Statistical Charts.}
As shown in Table~\ref{tab:chart_types}, basic statistical charts (bar, line, area) account for 46.9\% of all visualizations, consistent with the data-driven nature of analytical reports. The high prevalence of bar charts (26.74\%) reflects their versatility in comparative analysis, while the frequent use of line charts (15.12\%) indicates a focus on trend visualization.

\paragraph{Prominence of Process Visualization.}
Flowcharts rank third at 14.73\%, a notably high proportion for non-statistical charts. This suggests that the generated reports emphasize logical relationships and procedural explanations alongside raw data presentation. The combined relational and process chart category (25.6\%) demonstrates the system's capability to handle complex structural reasoning beyond simple data plotting.

\paragraph{Multimodal Type Diversity.}
Beyond basic statistical charts, the system generates a rich variety of specialized visualizations including geographic maps (6.98\%), structural diagrams (6.59\%), infographics (3.88\%), and matrices (3.10\%). This demonstrates the system's ability to select appropriate visual encodings for diverse analytical contexts—from spatial data (maps) to conceptual relationships (diagrams) to decision frameworks (matrices). The presence of 22 distinct chart types across 4 functional categories validates the system's multimodal reasoning capability.

\paragraph{Rendering Technology Distribution.}
The system employs a dual-technology stack: ECharts handles 81.9\% of visualizations (primarily data-driven charts and maps), while Mermaid manages 18.1\% (flowcharts and architectural diagrams). This division aligns well with each library's strengths—ECharts for quantitative visualization and Mermaid for declarative diagram syntax—resulting in efficient and appropriate technology allocation.

\paragraph{Coverage and Concentration.}
The type distribution exhibits a natural concentration pattern: the top 10 chart types cover 84.9\% of all visualizations, indicating a stable set of core visualization patterns. Simultaneously, the presence of specialized types (accounting for 15.1\% of charts) demonstrates the system's flexibility to adapt to domain-specific analytical needs. This balance between standardization and specialization reflects effective alignment between the system's multimodal generation capability and the diverse requirements of analytical report writing.

\section{OWID Dataset Construction}
\label{app:dataset}

We constructed our evaluation dataset from Our World in Data (OWID),%
\footnote{\url{https://ourworldindata.org}} a widely-cited platform for 
data-driven research reports. The construction involved three stages: 
web scraping, quality filtering, and format standardization.

\subsection{Data Collection}

We developed an automated web scraper to collect reports from OWID's 
publication archive (December 2016--September 2025). The scraper extracts 
complete report content (title, publication date, authors, main text, 
embedded visualizations) and implements politeness controls (1--2 second 
request delays, automatic retry mechanisms). This process collected 
399 reports across diverse topics including health, environment, economics, and social issues.

\subsection{Filtering}

To focus on substantive research reports and exclude announcements or atypical content, we applied the following criteria: Content length: 15,000--60,000 characters; Word count: $\geq$ 2,500 words; Visualizations: 3--15 images per report; Excluded keywords like ``Announcing'', ``Welcoming''.

The minimum requirements ensure sufficient content for meaningful evaluation, while maximum thresholds remove edge cases (e.g., comprehensive handbooks, image repositories). The visualization constraint focuses on typical research reports with substantive multimodal integration. Furthermore, we verified that the retained reports are free of sensitive personally identifiable information (PII). After filtering, 40 high-quality reports remained (10.04\% retention rate).

\subsection{Format Standardization}

Reports were standardized for evaluation use. HTML content was converted to Markdown format preserving document structure (headings, paragraphs, lists). Crucially, visualization references in text were mapped to their corresponding image files, maintaining the spatial and semantic relationships between text and visuals. This image-text alignment is essential for evaluating multimodal integration quality. Metadata (source, publication date, content statistics) was preserved for reproducibility.

\subsection{Dataset Statistics}

The compiled dataset comprises 40 reports, averaging 3,625 words and 7.9 visualizations per report. Reports span diverse topics with substantial multimodal content, providing a challenging testbed for automated report generation systems.

\section{Prompt}
In this section, we provide the evaluation prompts for our framework, including a template and metrics across five dimensions. These prompts were also used by human evaluators. Due to the large number of prompts required for individual agents and intermediate processes in CogGen, the system prompts will be released along with the code.

\onecolumn
\begin{AutoPrompt}{Prompt for Evaluation Template}
{query_section}{rubric}

---Separator: Below are two reports to be compared on the same dimension (including text and charts)---
{report1}

{report2}
---Separator: End of two report contents---

[Evaluation Task]
You need to **simultaneously** evaluate Report A (Model Report) and Report B (Reference Report) on the "{dimension_name}" dimension, and provide relative advantage judgment.

**Evaluation Method**:
1. Read both reports completely to form an overall quality impression
2. Please understand the intent of the user question and the purpose of the report, and consider whether the report's organization matches these intents and purposes
3. Determine which score range description (1-5 points) each report's overall performance is closer to
4. Score based on overall quality level

Please refer to the description of each score level (1-5 points) in the [Scoring Rubric] section of the rubric above, and determine:
- Which score range (integer between 1-5) Report A's overall performance on this dimension is closer to
- Which score range Report B's overall performance on this dimension is closer to
- Which one is overall better on this dimension, and what are the reasons

[Output Requirements]
Please output the comparison results in JSON format (do not include markdown code block markers):
{{
    "model_score": <integer from 1-5>,           // Score for Report A (Model Report)
    "reference_score": <integer from 1-5>,       // Score for Report B (Reference Report)
    "reasoning": "<Detailed comparison reasoning process, at least 150 words, comprehensively explaining the advantages and disadvantages of both reports and overall differences>",
    "evidence_model": ["<Specific evidence 1 from model report>", "<Specific evidence 2 from model report>"],
    "evidence_reference": ["<Specific evidence 1 from reference report>", "<Specific evidence 2 from reference report>"],
    "suggestions_model": ["<Specific improvement suggestion 1 for model report>", "<Specific improvement suggestion 2>"],
    "suggestions_reference": ["<Specific improvement suggestion 1 for reference report>", "<Specific improvement suggestion 2>"]
}}
\end{AutoPrompt}

\begin{AutoPrompt}{VISUAL-TEXT ALIGNMENT}
[Evaluation Dimension]Visual-Text Semantic Alignment

[Definition]Evaluate the formal integrity of visual-text integration, focusing on: (1) Reference clarity—whether text explicitly references figures (e.g., "as shown in Figure X", "the chart above illustrates"); (2) Transition smoothness—whether text naturally leads into figures and provides interpretation afterward; (3) Reading flow—whether visual-text switching feels natural and integrated into the narrative.

[Scoring Rubric]1-5 points

5 points: Seamless visual-text integration with excellent referencing and transitions
Text explicitly references each figure with clear pointers. Figures are naturally introduced by preceding text and followed by interpretation/discussion. The reading flow is smooth—figures feel like integral parts of the narrative, not insertions. Readers never wonder "why is this figure here?"

4 points: Good visual-text integration with clear referencing
Most figures have explicit text references. Transitions into and out of figures are generally smooth. Minor instances where a figure appears without clear introduction or follow-up discussion, but overall the integration is coherent.

3 points: Basic visual-text integration with inconsistent referencing
Some figures have explicit references, others appear without clear textual connection. Transitions are uneven—some figures flow naturally, others feel inserted. Readers can follow along but occasionally lose the connection between text and visuals.

2 points: Weak visual-text integration with poor referencing
Few explicit figure references. Figures often appear abruptly without introduction or interpretation. Text and visuals feel like separate elements rather than an integrated narrative. Readers must work to understand how figures relate to surrounding text.

1 point: Disconnected visual-text presentation
Almost no explicit figure references. Figures appear randomly with no textual connection. Text and visuals are essentially independent—removing figures would not disrupt text flow (indicating poor integration). Readers cannot understand the visual-text relationship.
\end{AutoPrompt}

\begin{AutoPrompt}{Multimodal Synergy}
[Evaluation Dimension]Multimodal Synergy

[Definition]Evaluate whether visuals and text work together to create understanding that exceeds what either could achieve alone. Key aspects: (1) Information increment—whether figures provide NEW information/perspectives beyond what text states (not just visual repetition of text content); (2) Complementary roles—whether text explains concepts while figures show data/relationships/patterns; (3) Synergistic effect—whether combining text and figures produces 1+1>2 understanding.

[Key Distinction]
- HIGH synergy: Figure shows data patterns/comparisons that text describes in words → reader gains both conceptual understanding AND visual evidence
- LOW synergy: Figure merely visualizes what text already fully explains → figure is decorative, removing it loses nothing
- Ask: "If I remove this figure, would the reader lose important information?" If NO, the figure lacks information increment.

[Scoring Rubric]1-5 points

5 points: Excellent synergy with strong information increment
Figures provide substantial information beyond text—showing patterns, comparisons, or relationships that text alone cannot efficiently convey. Text and figures have clear division of labor: text explains "why" and "what it means", figures show "what the data looks like". Removing figures would significantly reduce reader understanding. True 1+1>2 effect.

4 points: Good synergy with meaningful information increment
Most figures contribute information beyond text repetition. Text and figures generally complement each other. Some figures may slightly overlap with text content, but overall the combination enhances understanding noticeably.

3 points: Moderate synergy with limited information increment
Figures and text have some complementarity, but several figures mainly visualize what text already states. Information increment is inconsistent—some figures add value, others feel redundant. Removing some figures would not significantly impact understanding.

2 points: Weak synergy, figures largely repeat text
Most figures are visual restatements of text content without adding new information or perspectives. Little division of labor—text and figures say the same things in different formats. Figures feel like illustrations rather than information carriers.

1 point: No synergy, figures are purely decorative
Figures provide no information increment—they simply convert text statements into visual form. Removing all figures would not reduce information content. Text and figures are redundant rather than complementary. No 1+1>2 effect achieved.
\end{AutoPrompt}

\begin{AutoPrompt}{Information Organization}
[Evaluation Dimension]Information Organization Clarity

[Definition]Evaluate whether the report's structure, layout, and logical connections are clear. This includes: (1) Static structure—whether hierarchy is clear and complete; (2) Dynamic flow—whether sections have natural logical progression and smooth transitions. Clear organization can reduce the cognitive cost of visual search and comprehension.

[Scoring Rubric]1-5 points

5 points: Perfect structure with excellent logical flow
Report structure is complete, hierarchy is clear, sections progress in a natural logical order with smooth transitions between them. Readers can easily follow the reasoning from beginning to end.

4 points: Good structure with reasonable flow
Report structure is basically complete, hierarchy is basically clear, sections have reasonable logical order. Transitions between sections are adequate though not always seamless.

3 points: Average structure, weak logical flow
Report has basic structure, but logical progression between sections is weak. Some sections feel disconnected or the order seems arbitrary. Readers can understand individual sections but may struggle to see how they connect.

2 points: Messy structure, poor flow
Report has some structural elements but lacks clear logical progression. Sections appear randomly ordered, transitions are missing or abrupt. Readers have difficulty following the overall argument.

1 point: No organization, fragmented
Report has almost no structure, sections are like fragments randomly pieced together with no logical connection. Readers cannot understand the overall framework or how parts relate.
\end{AutoPrompt}

\begin{AutoPrompt}{Content Depth and Insight}
[Evaluation Dimension]Content Depth and Insight

[Definition]Evaluate whether the report provides appropriate depth across all important aspects of the topic. This dimension assesses: (1) Coverage completeness—whether all important facets of the topic are addressed (not just some aspects); (2) Depth balance—whether analysis depth is evenly distributed (not deep on some parts while shallow on others); (3) Analytical quality—whether the report provides mechanism explanations and causal reasoning, not just facts.

[Key Principle]
A well-planned report should comprehensively cover the topic with balanced depth across sections. Signs of poor planning include: some sections with rich analysis while others are superficial; important aspects of the topic missing entirely; depth that doesn't match section importance.

[Important Clarification]
- Depth ≠ Length: A long report with only surface-level facts is NOT deep; a concise report with insightful analysis IS deep
- Focus on analytical quality: mechanism explanations, causal reasoning, and meaningful insights—not word count or section length

[Scoring Rubric]1-5 points

5 points: Comprehensive coverage with balanced, high-quality depth
Report covers all important aspects of the topic thoroughly. Depth is well-balanced across sections—no section feels significantly more superficial or detailed than others relative to its importance. Each section provides meaningful analysis with mechanism explanations and causal reasoning. Readers gain complete understanding of the topic.

4 points: Good coverage with mostly balanced depth
Report covers most important aspects with good analytical depth. Depth distribution is reasonable, though minor imbalances exist (e.g., one section slightly more detailed than necessary, another slightly thin). Overall, readers get a solid understanding of the topic.

3 points: Incomplete coverage or unbalanced depth
Report has noticeable gaps: either some important aspects of the topic are missing, OR depth is clearly unbalanced (some sections have rich analysis while others are superficial lists). Readers understand parts of the topic well but lack insight into other parts.

2 points: Poor coverage or severely unbalanced depth
Report has significant coverage gaps—multiple important aspects are missing or barely touched. OR depth is severely unbalanced: detailed analysis on minor points while core aspects receive only surface treatment. Readers get fragmented, incomplete understanding.

1 point: Minimal coverage, shallow throughout
Report barely covers the topic—most important aspects are missing. What is covered lacks analytical depth (just facts, no mechanism explanations). Readers cannot form meaningful understanding of the topic.
\end{AutoPrompt}

\begin{AutoPrompt}{Content Relevance and Adaptation}
[Evaluation Dimension]Content Relevance and Adaptation

[Definition]Evaluate whether the report's content is relevant to its stated topic and appropriately structured as a comprehensive report. This dimension assesses: (1) Topic relevance—whether all substantive content relates to the report's subject matter; (2) Appropriate depth—whether the report provides sufficient context, background, and analysis expected of a quality report; (3) Non-redundancy—whether information is presented without excessive repetition across sections.

[Important Clarification]
- Background sections, methodology explanations, data source descriptions, and contextual information are LEGITIMATE parts of a quality report—they should NOT be penalized as "unnecessary content"
- Only penalize content that is truly OFF-TOPIC (unrelated to the subject) or EXCESSIVELY REPETITIVE (same points repeated verbatim multiple times)
- Meta-elements like citations, acknowledgments, and licensing notices are standard academic/journalistic conventions and should be IGNORED in this evaluation (neither rewarded nor penalized)

[Scoring Rubric]1-5 points

5 points: Highly relevant and well-structured report
All substantive content directly relates to the report topic. Background, analysis, and conclusions form a coherent whole. No off-topic digressions, no excessive repetition. The report covers the topic comprehensively without wandering.

4 points: Mostly relevant with minor issues
Report content is well-aligned with the topic. May have minor digressions or slight repetition, but these do not detract significantly from the overall coherence and relevance.

3 points: Moderately relevant with noticeable issues
Report addresses the topic but includes some off-topic sections or noticeable repetition of the same points across different sections. The core content is relevant but diluted by tangential material.

2 points: Poorly focused on the topic
Report has significant relevance problems: substantial off-topic content, major digressions from the subject matter, or excessive repetition that makes the report feel padded. Readers struggle to extract the relevant information.

1 point: Largely irrelevant or incoherent
Report barely addresses its stated topic. Dominated by off-topic content or so repetitive that little new information is conveyed. The report fails to deliver on its subject matter.
\end{AutoPrompt}

\end{document}